\documentstyle [12pt,twoside]{article}

\newcommand{\bq}{\begin{equation}}
\newcommand{\ba}{\begin{eqnarray}}
\newcommand{\eq}{\end{equation}}
\newcommand{\ea}{\end{eqnarray}}

\def\b{\beta}
\def\c{\raisebox{.4ex}{$\chi$}}
\def\d{\delta}
\def\e{\epsilon}
\def\f{\phi}

\def\h{\eta}

\def\j{\psi}

\def\m{\mu}
\def\n{\nu}
\def\o{\omega}
\def\p{\pi}
\def\q{\theta}
\def\r{\rho}
\def\s{\sigma}

\def\F{\Phi}
\def\G{\Gamma}

\def\O{\Omega}

\def\cl{{\cal L}}

\def\bo{{\raise.15ex\hbox{\large$\Box$}}}
\def\bob{{\lower.2ex\hbox{\large$\Box$}}}
\def\pa{\partial}
\def\de{\nabla}

\def\TH{{\raise.2ex\hbox{$\displaystyle \bigodot$}\mskip-4.7mu \llap H \;}}
\def\face{{\raise.2ex\hbox{$\displaystyle \bigodot$}\mskip-2.2mu \llap {$\ddot
        \smile$}}}

\def\Hat#1{\rlap{\kern.10em$\widehat{\phantom G}$}#1}
\def\HAt#1{\rlap{\kern.05em$\widehat{\phantom G}$}#1}

\def\cap#1{\rlap{\kern.1em$\widehat{\phantom{G\vrule height.8em}}$}#1{}}
\def\Cap#1{\rlap{\kern.05em$\widehat{\phantom{G\vrule height.8em}}$}#1{}}

\def\VEV#1{\left\langle #1\right\rangle}

\def\abs#1{\left| #1\right|}
\def\leftrightarrowfill{$\mathsurround=0pt \mathord\leftarrow \mkern-6mu
        \cleaders\hbox{$\mkern-2mu \mathord- \mkern-2mu$}\hfill
        \mkern-6mu \mathord\rightarrow$}
\def\overleftrightarrow#1{\vbox{\ialign{##\crcr
        \leftrightarrowfill\crcr\noalign{\kern-1pt\nointerlineskip}
        $\hfil\displaystyle{#1}\hfil$\crcr}}}

\def\frac#1#2{{\textstyle{#1\over\vphantom2\smash{\raise.20ex
        \hbox{$\scriptstyle{#2}$}}}}}

\def\nis{\nointerlineskip}
\def\Abar{\vbox{\nis\moveright.33em\vbox{
        \hrule width.35em height.04em}\nis\kern.05em\hbox{$A$}}{}}
\def\Dbar{\vbox{\nis\moveright.20em\vbox{
        \hrule width.50em height.04em}\nis\kern.05em\hbox{$D$}}{}}
\def\Gbar{\vbox{\nis\moveright.20em\vbox{
        \hrule width.50em height.04em}\nis\kern.05em\hbox{$G$}}{}}
\def\mbar{\vbox{\nis\moveright.15em\vbox{
        \hrule width.60em height.04em}\nis\kern.05em\hbox{$m$}}{}}
\def\Rbar{\vbox{\nis\moveright.20em\vbox{
        \hrule width.50em height.04em}\nis\kern.05em\hbox{$R$}}{}}
\def\Vbar{\vbox{\nis\moveright.05em\vbox{
        \hrule width.60em height.04em}\nis\kern.05em\hbox{$V$}}{}}
\def\Xbar{\vbox{\nis\moveright.20em\vbox{
        \hrule width.60em height.04em}\nis\kern.05em\hbox{$X$}}{}}
\def\thetabar{\vbox{\nis\moveright.15em\vbox{
        \hrule width.30em height.04em}\nis\kern.05em\hbox{$\theta$}}{}}
\def\Lambdabar{\vbox{\nis\moveright.25em\vbox{
        \hrule width.35em height.04em}\nis\kern.05em\hbox{${\mit\Lambda}$}}{}}
\def\Sigmabar{\vbox{\nis\moveright.25em\vbox{
        \hrule width.50em height.04em}\nis\kern.05em\hbox{${\mit\Sigma}$}}{}}
\def\phibar{\vbox{\nis\moveright.18em\vbox{
        \hrule width.40em height.04em}\nis\kern.05em\hbox{$\phi$}}{}}
\def\chibar{\vbox{\nis\moveright.12em\vbox{
        \hrule width.40em height.04em}\nis\kern.05em\hbox{$\chi$}}{}}
\def\psibar{\vbox{\nis\moveright.23em\vbox{
        \hrule width.40em height.04em}\nis\kern.05em\hbox{$\psi$}}{}}
\def\debar{\vbox{\nis\moveright.18em\vbox{
        \hrule width.35em height.04em}\nis\kern.05em\hbox{$\partial$}}{}}
\def\delbar{\vbox{\nis\moveright.10em\vbox{
        \hrule width.63em height.04em}\nis\kern.05em\hbox{$\nabla$}}{}}

\def\begintitle#1#2#3#4
        {\begin{titlepage}
         \centerline{#1 \hfill MdDP-PP-88-#2}
         \begin{center}\vglue .7in
         {\large\bf #3}\\].7in(
         {\bf #4}\\
         {\it Department of Physics and Astronomy}\\
         {\it University of Maryland, College Park, MD 20742}\\].7in(
         {\bf ABSTRACT}\\
         \end{center}
         \begin{quotation}}
\def\endtitle
         {\end{quotation}
          \end{titlepage}
          \newpage}

\begin{document}

\centerline{\large{\bf Stochastic Inflation: The Quantum Phase Space
Approach}\normalsize}

\vspace*{2cm}

\centerline{Salman Habib}
\centerline{\em Department of Physics}
\centerline{\em The University of British Columbia}
\centerline{\em 6224 Agriculture Road}
\centerline{\em Vancouver, B. C. V6T 1Z1}
\centerline{and}
\centerline{\em T-6, Theoretical Division$^{\dagger}$}
\centerline{\em Los Alamos National Laboratory}
\centerline{\em Los Alamos, NM 87545}

\vfill
\noindent e-mail: habib@eagle.lanl.gov\\
\noindent $^{\dagger}$Present Address\\
\newpage

\centerline{\bf Abstract}

In this paper a quantum mechanical phase space picture is constructed
for coarse-grained free quantum fields in an inflationary Universe. The
appropriate stochastic quantum Liouville equation is derived. Explicit
solutions for the phase space quantum distribution function are found
for the cases of power law and exponential expansions. The expectation
values of dynamical variables with respect to these solutions are
compared to the corresponding cutoff regularized field theoretic
results (we do not restrict ourselves only to $\VEV{\F^2}$). Fair
agreement is found provided the coarse-graining scale is kept within
certain limits. By focusing on the full phase space distribution
function rather than a reduced distribution it is shown that the
thermodynamic interpretation of the stochastic formalism faces several
difficulties (e.g., there is no fluctuation-dissipation theorem). The
coarse-graining does not guarantee an automatic classical limit as
quantum correlations turn out to be crucial in order to get results
consistent with standard quantum field theory. Therefore, the method
does {\em not} by itself constitute an explanation of the quantum to
classical transition in the early Universe. In particular, we argue
that the stochastic equations do not lead to decoherence.

\newpage
\centerline{\bf I. Introduction}

The paradigm of stochastic inflation, first introduced explicitly by
Starobinsky \cite{kn:aas}, has recently become popular as a means of
investigating various features of inflation. Some studies using this
formalism are those of density perturbations from inflation
\cite{kn:dens}, the very large scale structure of the Universe
\cite{kn:vls}, ``eternal inflation'' \cite{kn:eta}, power law
inflation \cite{kn:pli1}\cite{kn:pli2}, and speculations regarding the
relationship of this formalism to quantum cosmology \cite{kn:qc} (this
list is by no means exhaustive).

It must be admitted, however, that there is still no iron-clad
justification for the systematics of the method nor, for that matter,
a clear-cut interpretational scheme. The claim at issue is that
the infrared behavior of massless or small mass quantized scalar
fields in an inflationary Universe can be described in terms of a
real time classical random process. The source of the noise is taken
to be large scale quantum fluctuations which are continuously
generated in an inflationary Universe by red-shifting of the
ultraviolet sector. A key question here is: can these
quantum fluctuations be treated as being classical?

These fundamental issues have been considered previously for free
fields \cite{kn:tgp} however the situation for interacting fields is
not clear, and it is not obvious how far, if it all, any of the
present ``derivations'' are correct \cite{kn:blh}.
In this paper we leave aside for the moment the problem of interacting
fields and attempt a further clarification of the issues addressed in
Ref. \cite{kn:tgp}. To do so we will derive a phase space quantum master
equation for a quantum distribution function (the Wigner function) and
study its solutions. In our stochastic approach averages with respect
to this distribution function are supposed to reproduce quantum field
theoretic expectation values. A study of the solution itself is
supposed to enable one to judge the ``classicality'' of each physical
situation. The new method is distinct from the conventional approach
(where one takes as given a {\em classical} Langevin equation), and
enables the inclusion of crucial quantum correlations that have been
unjustifiably neglected in the past.

A subtle and important aspect of the Wigner distribution function is
the fact that essential quantum features are hidden in quantum
correlation ``cross-terms'' that disappear when one integrates over
any one of the phase space variables to produce a one-variable
(necessarily positive definite) distribution function. Such a reduced
distribution is {\em essentially useless} as a diagnostic tool for
studying quantum correlations in phase space (as will be seen
forcefully in this paper). Unfortunately it is on precisely such
objects that attention has been focused till now. Here, with the full
Wigner function at hand we will be able to go much further with regard
to clarifying the physics behind the stochastic approach. It has been
noticed previously \cite{kn:tgp} that the reduced distribution for a
massive scalar field in de Sitter space has, at late times, an
intriguing thermodynamic interpretation: it corresponds to a Boltzmann
distribution at the Gibbons-Hawking temperature. However, the full
distribution found in this paper does not have a thermal form even
though the reduced distribution is the same. This can be traced
directly to the fact that quantum correlations have not been
neglected; indeed they are every bit as important as the remaining
contributions. We will go more deeply into this question in Sec. V.

It has long been appreciated that the stochastic approach probes the
infrared sector of the relevant field theory. The length scale is set
by a certain parameter $\e$ which is usually taken to be small (i.e.,
attention is confined to length scales much larger than the time
dependent horizon length). Assuming inflation began at a finite time
in the past, one cannot take $\e$ to be arbitrarily small {\em
independent} of the time scale of interest: a small $\e$ is consistent
only with ``late times.'' It is also well known
that the quantum theory of free fields in an inflationary spacetime
has a nontrivial infrared sector and that a simple (though certainly
not rigorous) way to calculate expectation values of field variables
is to set an upper momentum cutoff at the Hubble scale (corresponding
to $\e \simeq 1$). The stochastic picture conflicts with these field
theoretic results unless $\e$ is small; this is due to the fact that
in the stochastic approach one focuses essentially on the zero mode
and attempts to include inhomogeneities only through a noise term.
Though this approximation appears to be quite drastic, we will
show that the stochastic calculations even for $\e\sim 1$ are never
too far from the naive field theoretic results. (Unlike some previous
work, our formalism does not restrict the value of $\e$.)

Previous work in stochastic inflation has concentrated mainly on the
quantity $\VEV{\F^2}$. In this paper we extend the method to compute
$\VEV{\F\p+\p\F}$ and $\VEV{\p^2}$. Such quantum averages would be
needed if one wished to compute the expectation value of the stress
tensor. (While for a small mass field, and an exponential inflation,
the dominant contribution to to $\VEV{T_{\m\n}}$ is from terms
$\propto\VEV{\F^2}$, it is important to check if the calculation of
the other terms is trustworthy.) Earlier approaches to stochastic
inflation implement approximations which led to incorrect values for
these quantities. Indeed precisely these approximations formed the
basis of some arguments about stochastic inflation leading to an
automatic classical limit for the quantum field theory. We will argue
against any such result in Sec. V.

An interesting (and somewhat uncomfortable) feature of the quantum
phase space distribution found in this paper is that, in some cases,
it depends quite strongly on $\e$ and is indeed singular in the limit
$\e\rightarrow 0$. Therefore, while it is true that $\VEV{\F^2}$ (as
well as the reduced distribution for the field alone) may be independent
of $\e$ for small values of $\e$, this parameter does not drop out of
the physics. A finite value of $\e$ is necessary for the distribution
function to exist; this is true even for a massive field in an
exponentially expanding Universe where to leading order all quadratic
phase space expectation values are independent of $\e$.

An important issue that seems to have received insufficient attention
in the stochastic inflation literature is a discussion of the role of
initial conditions. Massless theories in inflationary spacetimes
suffer from infrared divergences. Typically these divergences are
``fixed'' by assuming that inflation began a finite time in the past
and thereby modifying the infrared structure of the quantum state of
the field. Expectation values then have two contributions: one each
from the pre-inflationary and inflationary sectors. One can show that
the pre-inflationary contribution falls rapidly with time and can always
be neglected compared to the inflationary one (see the Appendix). In
the stochastic paradigm there are also two contributions to
expectation values: a systematic piece arising from the dynamical
evolution of an initial condition and a stochastic piece due entirely
to the noise source. We will show that while at late times, and for
arbitrary initial conditions, the second piece always dominates the
first, this is {\em not} true at early times. While the matching of
the quantum states in the pre-inflationary and inflationary regimes at
the onset of inflation can also provide an initial condition for the
stochastic method, the time dependences of the systematic contribution
in the stochastic method do not always match the time dependences of the
pre-inflationary contribution in the field theoretic calculation. This
fact coupled with the small $\e$ restriction might limit the
application of stochastic techniques in accurately studying the onset of
inflation. The use of the method for studying phase transitions in the
early Universe should also be approached with some caution
\cite{kn:cal}. (Of course all this is not a serious problem if one is
only interested in late time results.)

The attempt in this paper is to push the formalism of stochastic
inflation as hard as possible in simple examples: we find that some of
the appealing original results no longer appear as compelling as at
first sight. However, it is still a remarkable fact that a simple
stochastic model suffices to (almost) correctly calculate field
theoretic expectation values and further that the essentially
nonstationary phase space distribution nevertheless yields a thermal
(or ``random walk'') distribution for the reduced distribution
function. Whether this has a deep significance is unfortunately not
clear.

The organization of the paper is as follows: In Sec. II we derive the
appropriate stochastic quantum Liouville equation for the
coarse-grained field using the phase space formulation of quantum
mechanics and obtain the general solution. In Sec. III we apply these
results to the case of an inflationary expansion; power law expansions
are dealt with in Sec. IV. The existence of classical stochastic
interpretations is discussed in Sec. V via a study of the solutions of
the quantum Liouville equation. We conclude with Sec. VI where the
results are reviewed and future directions for research are suggested.
The quantum field theoretic derivations of the results obtained via the
stochastic approach are given in an Appendix.

\newpage

\centerline{\bf II. The Stochastic Quantum Liouville Equation}

In this section we will set up a formalism to study the evolution of
coarse-grained free scalar fields in a spatially flat inflationary
Friedmann-Robertson-Walker Universe. We will work under the ``test
field'' assumption, i.e., the contribution to the stress tensor from
the field is taken to be small compared to that of the matter driving
the expansion. All of our results will therefore not be applicable to
an inflaton field but some may indeed be extended to that case.

The line element for the spacetime is
\begin{eqnarray}
ds^2&=&-dt^2+a(t)^2d\vec{x}\cdot d\vec{x}    \label{aone}\\
&=&S(\h)^2\left(-d\h^2+d\vec{x}\cdot d\vec{x}\right), \label{atwo}
\end{eqnarray}
where in terms of the cosmic time $t$, the conformal time
\begin{equation}
\h=\int^t {dt'\over a(t')}.          \label{athree}
\end{equation}

A massive minimally coupled scalar field has the Lagrangian
\begin{equation}
{\cal L}(\F,\F_{,\m})=-{1\over
2}\sqrt{-g}\left(g^{\m\n}\F_{,\m}\F_{,\n} + m^2\F^2\right), \label{afour}
\end{equation}
which, with the metric choice (\ref{atwo}), reduces to
\begin{equation}
{\cal L}(\F,\F_{,\m})=-{1\over
2}\left(-S^2\dot{\F}^2+S^2\F_{,i}\F_{,i}+S^4m^2\F^2 \right). \label{afive}
\end{equation}
The overdot represents differentiation with respect to the
conformal time. In terms of the ``conformal field,'' defined via the
time dependent canonical transformation,
\begin{equation}
\c\equiv S\F,              \label{asix}
\end{equation}
and modulo an integration by parts, the Lagrangian (\ref{afive}) becomes
\begin{equation}
{\cal L}(\c,\c_{,\m})=-{1\over
2}\left[-\dot{\c}^2+\c_{,i}\c_{,i}+\left(S^2m^2-{\ddot{S}\over
S}\right)\c^2\right].             \label{aseven}
\end{equation}
One advantage of working with the conformal field and the Lagrangian
(\ref{aseven}) is that the equation of motion for the field
\begin{equation}
\ddot{\c}-\de^2\c+\left(S^2m^2-{\ddot{S}\over S}\right)\c=0 \label{aeight}
\end{equation}
does not have the first derivative in time ``Hubble damping'' term
found in the equation of motion for the original field
\begin{equation}
\ddot{\F}+2{\dot{S}\over S}\dot{\F}-\de^2\F+S^2m^2\F=0,  \label{anine}
\end{equation}
and that the canonical momentum
\begin{equation}
\p_{\tiny \c}\equiv{\pa\cl\over\pa\dot{\c}}=\dot{\c}    \label{aten}
\end{equation}
is of the usual flat space form.

The Hamiltonian corresponding to (\ref{aseven}) is
\begin{equation}
H(\c,\p_{\tiny \c})={1\over 2}\int
d\vec{x}\left[\p_{\tiny \c}^{~2}+\c_{,i}\c_{,i}+\left(S^2m^2-{\ddot{S}\over
S}\right)\c^2\right]            \label{aeleven}
\end{equation}
which is that for a free field in flat spacetime with a time dependent
mass. The Hamiltonian equations of motion are
\begin{eqnarray}
\dot{\c}&=&{\d H\over\d\p_{\tiny \c}}=\p_{\tiny \c}, \label{aelevena}\\
\dot{\p}_{\tiny \c}&=&-{\d H\over\d\c}=\de^2\c-\left(S^2m^2-{\ddot{S}\over
S}\right).     \label{aelevenb}
\end{eqnarray}

The form of the Hamiltonian (\ref{aeleven}), though ``canonical,'' is
hardly unique; if we had worked with some other choice of time and
field it would have been ``natural'' to consider a different
description in terms of a different Hamiltonian. At the level of a
classical treatment, and even at the level of quantum dynamics, this
difference is largely irrelevant. However, there is an aspect of the
quantum treatment where such a difference does indeed matter: The two
Hamiltonian descriptions will, upon canonical quantization, lead to
inequivalent descriptions in terms of different vacua. Furthermore,
there are well known difficulties if one chooses to select the time
dependent ground state of the Hamiltonian as the ``instantaneous
diagonalization'' vacuum \cite{kn:sf}. These problems will be of no
concern to us as in our case the choice of quantum state will be an
independently defined adiabatic vacuum (details will be given later).
(We note in passing that for spatially flat FRW models, Weiss has
shown \cite{kn:nw} that with the specific Hamiltonian (\ref{aeleven})
the instantaneous diagonalization approach can be made consistent
with a ``mode quantization'' for certain special choices of the
latter.)

In our case there is a conceptually important consequence of
(\ref{aeleven}) being the chosen Hamiltonian. In the stochastic
inflation literature there appears to be a tendency of interpreting
the Hubble damping term in (\ref{anine}) as being of a truly
dissipative nature. This runs the risk of repeating an old error in
quantum mechanics: the confusion of a time dependent mass with true
damping \cite{kn:damp}. Working with (\ref{aeleven}) and the
associated equation of motion (\ref{aeight}) manifestly eliminates the
possibility of such misinterpretations.

The scalar field is now quantized in the standard manner
\cite{kn:qft}; first we introduce the modes
\begin{eqnarray}
\c_{\vec{k}}(\vec{x},\h)&=&S(\h)\F_{\vec{k}}(\vec{x},\h) \label{atwelve}\\
&=&{\hbox{e}^{i\vec{k}\cdot\vec{x}}\over(2\p)^{3/2}}\c_k(\h) \label{athirteen}
\end{eqnarray}
where $\c_k(\h)$ is a solution of
\begin{equation}
\ddot{\c}_k+\o_k^{~2}\c_k=0  \label{afourteen}
\end{equation}
with $\o_k$ the oscillator ``frequency,'' defined via
\begin{equation}
\o_k^{~2}\equiv k^2+\left(S^2m^2-{\ddot{S}\over S}\right). \label{afourteenb}
\end{equation}
The annihilation and creation operators with respect to these modes,
which satisfy the commutation relations,
\begin{eqnarray}\label{afourteenc}
[\hat{a}_{\vec{k}},\hat{a}_{\vec{k}'}^{\dagger}]&=&\d(\vec{k}-\vec{k}'),\\
\label{afourteend}
[\hat{a}_{\vec{k}},\hat{a}_{\vec{k}'}]&=& 0
\end{eqnarray}
are then used to build the field operator
\begin{eqnarray}
{\hat{\c}}(\vec{x},\h)&=&S(\h)\hat{\F}(\vec{x},\h)     \label{afifteen}\\
&=&\int d\vec{k}~\left[\hat{a}_{\vec{k}}\c_{\vec{k}}(\vec{x},\h)
+\hat{a}_{\vec{k}}^{\dagger}\c_{\vec{k}}^*(\vec{x},\h)\right].
\label{asixteen}
\end{eqnarray}
As is well known the annihilation and creation operators are not
uniquely specified by (\ref{afourteenc}) and (\ref{afourteend}).
Further restrictions are needed to fix these operators and thereby
to uniquely specify the ``vacuum'' state annihilated by
$\hat{a}_{\vec{k}}$. We will turn to these issues shortly.

A seemingly generic feature of inflationary spacetimes is the
``destabilization'' of massless scalar fields
\cite{kn:destab}\cite{kn:despl} due in part to infrared divergences
\cite{kn:fp}. To render the quantum state infrared finite one assumes
a benign Robertson-Walker expansion in which there is no infrared
divergent adiabatic vacuum prior to inflation (a radiation dominated
Universe, for example). The quantum state in the inflationary phase is
matched to an infrared finite quantum state (e.g., the conformal
vacuum) at the time when inflation takes over from the previous epoch.
It is then possible to show that the new state is always free from
infrared divergences \cite{kn:fp}. The key result, however, is that
the expectation value $\VEV{\F^2}$ in the infrared finite (and
ultraviolet regulated) state starts to grow at the onset of
inflation; for power law inflation $\VEV{\F^2}$ grows to an asymptotic
constant value, whereas in the case of an exponential expansion it
grows linearly with cosmic time without any upper limit. These
otherwise puzzling results have a natural interpretation within the
framework of stochastic inflation \cite{kn:tgp}.

Crudely speaking, ``destabilization'' occurs when, with $m=0$, the
$\o_k^{~2}$ term in (\ref{afourteen}) goes negative with the passage
of time, at ever higher values of $k$. Modes at long wavelengths
behave as amplitudes for upside down harmonic oscillators (with time
dependent ``frequencies'') and due to the inflationary expansion there
is a continuous flow of short wavelength modes into this unstable
infrared sector. Therefore we focus attention on the long wavelength
modes, i.e., those with $k^2 < \ddot{S}/S$ by defining the
coarse-grained quantum field
\begin{equation}
\hat{\c}_L(\vec{x},\h)\equiv\int d\vec{k}~\q(k_S-k)
\left[\hat{a}_{\vec{k}}\c_{\vec{k}}(\vec{x},\h)+
\hat{a}_{\vec{k}}^{\dagger}\c_{\vec{k}}^*(\vec{x},\h)\right]
\label{aseventeen}
\end{equation}
and the corresponding coarse-grained momentum
\begin{equation}
\hat{\p}_L(\vec{x},\h)\equiv\int
d\vec{k}~\q(k_S-k)\left[\hat{a}_{\vec{k}}\dot{\c}_{\vec{k}}(\vec{x},\h)+
\hat{a}_{\vec{k}}^{\dagger}\dot{\c}_{\vec{k}}^*(\vec{x},\h)\right].
\label{aeighteen}
\end{equation}
For the moment we will leave the upper cutoff $k_S$ unspecified
beyond the fact that it is set by $\ddot{S}/S$ (however, it is
important to remember that this cutoff is time dependent). The
corresponding short wavelength fields $\hat{\c}_S$ and $\hat{\p}_S$
are defined by
\begin{eqnarray}
\hat{\c}&=&\hat{\c}_L+\hat{\c}_S,               \label{anineteen}\\
\hat{\p}_{\tiny \c}&=&\hat{\p}_L+\hat{\p}_S.              \label{atwenty}
\end{eqnarray}

The Heisenberg operators $\hat{\c}$ and $\hat{\p}_{\tiny \c}$ satisfy the
classical Hamiltonian equations of motion. Substituting
(\ref{anineteen}) and (\ref{atwenty}) in (\ref{aelevena}) and
(\ref{aelevenb}) we find
\ba
\dot{\hat{\c}}_L&=&\hat{\p}_L+\hat{F}_1^c,     \label{atwentyone}\\
\dot{\hat{\p}}_L&=&\de^2\hat{\c}_L-\left(S^2m^2-{\ddot{S}\over
S}\right)\hat{\c}_L+\hat{F}_2^c,
                            \label{atwentytwo}
\ea
where
\ba
\hat{F}_1^c(\vec{x},\h)&\equiv&\dot{k}_S\int d\vec{k}~\d(k-k_S)
\left[\hat{a}_{\vec{k}}\c_{\vec{k}}(\vec{x},\h)+
\hat{a}_{\vec{k}}^{\dagger}\c_{\vec{k}}^*(\vec{x},\h)\right],\label{atwentythree}\\
\hat{F}_2^c(\vec{x},\h)&\equiv&\dot{k}_S\int d\vec{k}~\d(k-k_S)
\left[\hat{a}_{\vec{k}}\dot{\c}_{\vec{k}}(\vec{x},\h)+
\hat{a}_{\vec{k}}^{\dagger}\dot{\c}_{\vec{k}}^*(\vec{x},\h)\right].
\label{atwentyfour}
\ea
The new terms $\hat{F}_1^c$ and $\hat{F}_2^c$ arise simply because
$k_S$ is time dependent. These terms represent the inflow of short
wavelength modes into the infrared ``condensate.'' (It is important to
note that this contribution exists even for free fields.)

In order to proceed further we have to decide which quantum state the
field is in during the inflationary phase. It is known that the
adiabatic vacuum suffers from infrared divergences; to produce states
free of such divergences one usually modifies the long wavelength mode
structure of the quantum state (i.e., long compared to the time
dependent horizon length at the onset of inflation) but leaves the short
wavelength structure the same. This implies that the quantum state for
computing expectation values of $\hat{F}_1^c$ and $\hat{F}_2^c$, and
of various powers of these operators, is the adiabatic vacuum. Other
choices are possible when describing different physical situations,
for example, thermal states have been considered in Ref. \cite{kn:pli2}
and more general vacuum states in Ref. \cite{kn:tgp}.

Eventually we will deal specifically with an exponential expansion and
with power law inflation (i.e., where the radius of the Universe goes
as a power, greater than one, of the cosmic time). For such a
Robertson-Walker Universe we will assume the quantum state for the
short wavelength modes to be the adiabatic vacuum (which reduces to
the Bunch-Davies vacuum \cite{kn:bad} for de Sitter space). For the
moment, though, all that is relevant is that the chosen state be
annihilated by the operator $\hat{a}_{\vec{k}}$ of (\ref{asixteen}),
so that,
\ba
\VEV{\hat{F}_1^c(\vec{x},\h)}&=&0,            \label{atwentyfive}\\
\VEV{\hat{F}_2^c(\vec{x},\h)}&=&0.            \label{atwentysix}
\ea
It is also straightforward to compute that
\bq
\VEV{\hat{F}_i^c(\vec{x}_1,\h_1)\hat{F}_j^c(\vec{x}_2,\h_2)}=
2B_{ij}(\vec{x}_1,\vec{x}_2,\h_1)\d(\h_1-\h_2),  \label{atwentyseven}
\eq
where, with $R\equiv\abs{\vec{x}_1-\vec{x}_2}$,
\ba
B_{11}(\vec{x}_1,\vec{x}_2,\h_1)&=&{1\over 4\p^2}k_S^{~2}\abs{\dot{k}_S}
{\sin k_SR\over k_SR}\abs{\c_{k_S}(\h_1)}^2,  \label{atwentyeight}\\
B_{12}(\vec{x}_1,\vec{x}_2,\h_1)&=&{1\over 4\p^2}k_S^{~2}\abs{\dot{k}_S}
{\sin k_SR\over k_SR}\c_{k_S}(\h_1)\dot{\c}_{k_S}^*(\h_1),
\label{atwentynine}\\
B_{21}(\vec{x}_1,\vec{x}_2,\h_1)&=&{1\over 4\p^2}k_S^{~2}\abs{\dot{k}_S}
{\sin k_SR\over k_SR}\dot{\c}_{k_S}(\h_1)\c_{k_S}^*(\h_1), \label{athirty}\\
B_{22}(\vec{x}_1,\vec{x}_2,\h_1)&=&{1\over 4\p^2}k_S^{~2}\abs{\dot{k}_S}
{\sin k_SR\over k_SR}\abs{\dot{\c}_{k_S}(\h_1)}^2. \label{athirtyone}
\ea
It is at this point that a stochastic interpretation suggests itself.
The quantum expectation value may be regarded as an averaging bracket
for the white (albeit nonstationary) ``noise'' operators $\hat{F}_1^c$ and
$\hat{F}_2^c$. Since we are dealing with a free theory it is trivial to
verify that the higher moments of these operators are those
appropriate for Gaussian noise. The fact that the noise is white stems
from the theta function cutoff in momentum space. Other cutoffs are
certainly acceptable, however, they will lead to the noises being
colored and unnecessarily complicate the derivation of the phase space
picture. We emphasize that physical results do not depend strongly on
this choice.

The ``diffusion matrix'' $B_{ij}$ has two curious features: it is
complex (albeit Hermitian) and singular. We will show later that as
far as the stochastic quantum Liouville equation is concerned what is
really relevant is $B_{ij}+B_{ji}$ which not only is necessarily real
but also has a nonzero determinant. The complex nature of $B_{ij}$ is
essential for $B_{ij}+B_{ji}$ to be nonsingular. It is important to be
cautious when implementing approximations for the diffusion matrix and
not to prematurely throw out the essential imaginary pieces. Finally,
the fact that $B_{12}$ and $B_{21}$ are complex implies that these
cross-correlations cannot be understood on a purely classical basis.

According to the conventional stochastic interpretation we should view
(\ref{atwentyone}) and (\ref{atwentytwo}) as Langevin equations for the
{\em classical} stochastic variables $\c_L$ and $\p_L$ with $F_1^c$ and
$F_2^c$ viewed as {\em classical} noises \cite{kn:sas}. However, it is
not at all obvious why this should be true. The Langevin equations
(\ref{atwentyone}) and (\ref{atwentytwo}) are {\em operator} equations
and we must have further information about the quantum state of the
system before any classical interpretation can be accepted.
Furthermore, the noise operators do not commute:
\bq
\left[\hat{F}_1^c(\vec{x}_1,\h_1),\hat{F}_2^c(\vec{x}_2,\h_2)\right]=
{i\over 2\p^2}k_S^{~2}\abs{\dot{k}_S}{\sin k_SR\over
k_SR}\d(\h_1-\h_2). \label{thirtytwo}
\eq
In principle the noises are certainly not classical: we will go on to
show that neglecting the quantum correlations buried in the noises
produces results conflicting with standard field theory. (The question
of the quantum state and the quantum nature of the coarse-grained
fields has been taken up in more detail in Ref. \cite{kn:tgp} where it has
been pointed out that the coarse-graining by itself does not lead to a
set of classical equations.) Here we follow a different path by
deriving a {\em quantum} stochastic Liouville equation that
incorporates, at least to some extent, the correlations between the
noises.

The fact that $\hat{F}_1^c$ and $\hat{F}_2^c$ do not commute implies
that (\ref{atwentyone}) and (\ref{atwentytwo}) should be treated as
two separate Langevin equations. Strictly speaking it is not valid to
substitute (\ref{atwentyone}) in (\ref{atwentytwo}) and treat the
resulting equation as a stochastic equation second order in time. This
will lead to wrong answers for averages involving $\hat{\p}_L$. The
approach we will follow avoids this pitfall.

The Hamiltonian equations of motion (\ref{atwentyone}) and
(\ref{atwentytwo}) are exact as no approximations have been made so
far. The first approximation we make is to drop the spatial derivative
term in (\ref{atwentytwo}); this is because we will be interested only
in the behavior of the quantum field at ``large'' scales. The
coarse-graining will be implemented in the sense of a temporal
ensemble, i.e., we focus attention on one spatially fixed
coarse-grained domain and consider the evolution of the coarse-grained
quantum field defined on that domain. The spatial coarse-graining and
this interpretation imply that all two-point objects be evaluated with
the spatial separation between the points being much less than the
coarse-graining scale, i.e., $R\ll 2\p k_S^{-1}$. With this limit in
place and with the neglect of spatial derivatives, the quantum Langevin
equations are
\ba
\dot{\hat{\c}}_L&=&\hat{\p}_L+\hat{F}_1^c, \label{athirtythree}\\
\dot{\hat{\p}}_L&=&-\left(S^2m^2-{\ddot{S}\over
S}\right)\hat{\c}_L+\hat{F}_2^c,         \label{athirtyfour}
\ea
where
\bq
\VEV{\hat{F}_i^c(\h_1)\hat{F}_j^c(\h_2)}\simeq
2B_{ij}(\h_1)\d(\h_1-\h_2),  \label{atwentysevenb}
\eq
and
\ba
B_{11}(\h_1)&\simeq&{1\over 4\p^2}k_S^{~2}\abs{\dot{k}_S}
\abs{\c_{k_S}(\h_1)}^2,  \label{athirtyfive}\\
B_{12}(\h_1)&\simeq&{1\over 4\p^2}k_S^{~2}\abs{\dot{k}_S}
\c_{k_S}(\h_1)\dot{\c}_{k_S}^*(\h_1), \label{athirtysix}\\
B_{21}(\h_1)&\simeq&{1\over 4\p^2}k_S^{~2}\abs{\dot{k}_S}
\dot{\c}_{k_S}(\h_1)\c_{k_S}^*(\h_1), \label{athirtyseven}\\
B_{22}(\h_1)&\simeq&{1\over 4\p^2}k_S^{~2}\abs{\dot{k}_S}
\abs{\dot{\c}_{k_S}(\h_1)}^2. \label{athirtyeight}
\ea
Spatial variations within one coarse-grained domain cannot be sampled
by the coarse-grained field; this accounts for the fact that there are
no terms reflecting such a dependence in
(\ref{athirtythree})--(\ref{athirtyeight}).

The dynamical equations (\ref{athirtythree}) and
(\ref{athirtyfour}) can just as well be obtained from the stochastic
Hamiltonian
\bq
H(\c_L,\p_L)={1\over
2}\p_L^{~2}+{1\over 2}\o^2(\h)\c_L^{~2}+F_1^c\p_L-F_2^c\c_L,
\label{athirtynine}
\eq
where the time dependent ``frequency,''
\bq
\o^2\equiv S^2m^2-{\ddot{S}\over S}.  \label{athirtynineb}
\eq
The coarse-grained field is now viewed as the coordinate variable in
the one-dimensional quantum mechanical problem specified by
(\ref{athirtynine}). The terms containing $F_1^c$ and $F_2^c$
are taken to represent stochastic external perturbations with
correlations specified by (\ref{atwentysevenb})--(\ref{athirtyeight}).
(The Hamiltonian (\ref{athirtynine}) is a time dependent
generalization of the randomly forced oscillator considered previously
in a different context by Merzbacher \cite{kn:merz}.) The idea now is
to study the one-dimensional quantum mechanical problem instead of the
original field theory. It is important to note that for a quantum
analysis we cannot just use the equations of motion
(\ref{athirtythree}) and (\ref{athirtyfour}); a Hamiltonian is
{\em necessary}. On the other hand were we only interested in a
classical analysis, the equations of motion would suffice. (A
discussion of this point is given in Ref. \cite{kn:shst}.)

Before proceeding further some cautionary remarks are in order. First,
while the above assumption is an improvement on previous work to the
extent that we are not assuming the system to be classical, it still
does not constitute a well controlled approximation scheme. In
particular, the Hamiltonian (\ref{athirtynine}) has been written down
simply by fiat. Nevertheless, to see whether the results and insights
obtained using this approach are persuasive, our attitude will be to
take the formalism as it stands and proceed as far as possible without
any further assumptions. Second, there is a coordinate dependence
inherent in the phase space formalism we will be employing shortly:
the distribution function is not invariant under canonical
transformations (this feature is generic to quantum mechanics and is
not specific to our problem). We will return to these problems in more
detail later on.

The quantum system is completely described by its density matrix,
which written in the coordinate representation,
\bq
\r(\c_L,\c_L')=\sum_j W_j\j_j(\c_L)\j_j^*(\c_L'), \label{aforty}
\eq
obeys the quantum Liouville equation
\bq
i\dot{\r}(\c_L,\c_L')=[H(\c_L)-H^*(\c_L')]\r(\c_L,\c_L'). \label{afortyone}
\eq
The passage to a quantum phase space is now made via the Wigner
transform \cite{kn:wig} of the density matrix:
\bq
f_W(X_L,p_L)=\int {dx_L\over
2\p}~\hbox{e}^{ip_Lx_L}\r(X_L+x_L/2,X_L-x_L/2),   \label{afortytwo}
\eq
where the new variables
\ba
X_L&=&\left(\c_L+\c_L'\right)/2,        \label{afortythree}\\
x_L&=&\c_L-\c_L'.                       \label{afortyfour}
\ea
The Wigner function $f_W(X_L,p_L)$ is always real and properly
normalized over phase space (for bounded systems), moreover it is
square integrable (a property not shared in general by classical
distribution functions):
\ba
\int dX_Ldp_L~f_W(X_L,p_L)&=&1,          \label{afortyfive}\\
\int dX_Ldp_L~f_W^{~2}(X_L,p_L)&\leq& {1\over 2\p},  \label{afortysix}
\ea
where in the second expression the equality holds for pure states.
Quantum expectation values for functions of $\hat{\c}_L$ and
$\hat{\p}_L$ alone are given correctly as phase space averages
with respect to the Wigner function, as for example,
\bq
\VEV{h(\hat{\p}_L)}=\int dX_Ldp_L~h(p_L)f_W(X_L,p_L) \label{afortyseven}
\eq
but not for mixed operators such as $\hat{\c}_L^{~2}\hat{\p}_L^{~2}$.
This is related to the ordering problem in quantum mechanics; the
Wigner formalism is associated with Weyl's rule for the ordering of
operators \cite{kn:coh}. A further obstacle to the literal
interpretation of a Wigner function as a true distribution function
over a classical phase space is the fact that in general it is not
positive definite. (Fortunately we will not encounter the ordering
problem nor the lack of positivity in our example.) More on the Wigner
function can be found in the reviews of Hillery {\em et al}
\cite{kn:hill} and Narcowich \cite{kn:narc}.

The Wigner transform of the quantum Liouville equation
(\ref{afortyone}) yields
\bq
{\pa\over\pa\h} f_W(X_L,p_L;\h)=-L_0 f_W(X_L,p_L;\h)-L_S
f_W(X_L,p_L;\h),          \label{afortyeight}
\eq
where the Liouville operator has been written as the sum of a
systematic piece
\bq
L_0=p_L{\pa\over\pa X_L}-\o^2X_L{\pa\over\pa p_L} \label{afortynine}
\eq
and a stochastic piece
\ba
L_S&=&F_1^c{\pa\over\pa X_L}+F_2^c{\pa\over\pa p_L} \label{afifty}\\
&\equiv&F^c_i{\pa\over\pa z_i},~~~~~~~~~(z_1\equiv X_L,z_2\equiv p_L).
\label{afiftyone}
\ea
We now implement the strategy of Kubo \cite{kn:kubo} in order to
obtain a simple derivation of the stochastic quantum Liouville
equation (cases more complicated than the one considered here are
treated elsewhere \cite{kn:shst}). To begin, we focus attention on the
dynamical effect of $L_S$ by shifting to the interaction picture:
\bq
 f_W(X_L,p_L;\h)=\hbox{e}^{-L_0\h} \s(X_L,p_L;\h).  \label{afiftytwo}
\eq
In terms of $\s(X_L,p_L;\h)$ the Liouville equation
(\ref{afortyeight}) becomes
\ba
{\pa\over\pa\h}\s(X_L,p_L;\h)&=&-\hbox{e}^{L_0\h}L_S
\hbox{e}^{-L_0\h}\s(X_L,p_L;\h)      \label{afiftythree}\\
&\equiv& \O(\h)\s(X_L,p_L;\h).       \label{afiftyfour}
\ea
This equation has the formal time ordered exponential solution
\ba
\s(X_L,p_L;\h)&=&\left[1+\int_{\h_0}^{\h}d\h_1~\O(\h_1)+
\int_{\h_0}^{\h}d\h_1\int_{\h_0}^{\h_1}d\h_2~\O(\h_1)\O(\h_2)+
\cdots\right]\s(X_L,p_L;\h_0)       \nonumber\\
&=&\left[\exp\left(\int_{\h_0}^{\h}d\h'~\O(\h')\right)\right]_T
\s(X_L,p_L;\h_0), \label{afiftysix}
\ea
where the initial value $\s(X_L,p_L;\h_0)$ is specified at some
initial time $\h_0$. All the noise terms come multiplied together in
each term of the series. If we take the average over noise of
(\ref{afiftysix}), these terms will either be zero, or will produce
delta functions. It is easy to see that only the quadratic product of
noise terms needs to be computed, this following from the Gaussian
nature of the noises. With $\VEV{~}_N$ denoting an average over noise,
we find,
\bq
\VEV{\O(\h_1)\O(\h_2)}_N=2B_{ij}(\h_1)\d(\h_1-\h_2)
\left[\hbox{e}^{L_0\h_1}{\pa^2\over\pa z_i\pa
z_j}\hbox{e}^{-L_0\h_1}\right].            \label{afiftyseven}
\eq
The noise averaged version of the time ordered exponential solution
(\ref{afiftysix}) then turns out to be
\bq
\VEV{\s(X_L,p_L;\h)}_N=\s(X_L,p_L;\h_0)+\int_{\h_0}^{\h}d\h_1~
B_{ij}(\h_1)\left[\hbox{e}^{L_0\h_1}{\pa^2\over\pa z_i\pa
z_j}\hbox{e}^{-L_0\h_1}\right] \VEV{\s(X_L,p_L;\h_1)}_N, \label{afiftyeight}
\eq
which may be immediately differentiated to yield
\bq
{\pa
\over\pa\h}\VEV{\s(X_L,p_L;\h)}_N=B_{ij}(\h)
\left[\hbox{e}^{L_0\h}{\pa^2\over\pa z_i\pa
z_j}\hbox{e}^{-L_0\h}\right]\VEV{\s(X_L,p_L;\h)}_N.  \label{afiftynine}
\eq
We recall that the transformation to the interaction picture involved
only $L_0$ which is of course unaffected by averages over the noise.
Therefore, there is no difficulty in writing (\ref{afiftynine}) in
terms of the original distribution function:
\bq
{\pa\over\pa\h}\VEV{f_W(X_L,p_L;\h)}_N=-L_0\VEV{f_W(X_L,p_L;\h)}_N+
B_{ij}(\h){\pa^2\over\pa z_i\pa z_j}\VEV{f_W(X_L,p_L;\h)}_N. \label{asixtya}
\eq
This is the required stochastic quantum Liouville equation and, as is
obvious, it has the standard Fokker-Planck form. Alternatively,
(\ref{asixtya}) may be written in a more convenient form in terms of
the explicitly symmetrized diffusion matrix $D_{ij}=(B_{ij}+B_{ji})$
as
\bq
{\pa\over\pa\h}\VEV{f_W(X_L,p_L;\h)}_N=-L_0\VEV{f_W(X_L,p_L;\h)}_N+
{1\over 2}D_{ij}(\h){\pa^2\over\pa z_i\pa z_j}\VEV{f_W(X_L,p_L;\h)}_N.
\label{asixty}
\eq
The stochastic equation (\ref{asixty}) and the nature of its derivation
merit a few clarificationary remarks. First, it is not necessary to
begin with the Wigner formalism; we can just as well employ the
density matrix (either by following the procedure used here or the
influence functional approach \cite{kn:fv}). The stochastic equation
for the density matrix can then be converted to one for the Wigner
function by implementing the ``twisted product'' \cite{kn:narc}.
Second, as only the case of free fields is treated here, the
Hamiltonian is at most quadratic in the dynamical variables. This is
why (\ref{asixty}) is of the standard classical Fokker-Planck form;
such a simplification does not obtain in general \cite{kn:shst}. Of
course, even if the form of (\ref{asixty}) is classical, this does
{\em not} imply that all solutions be classical distribution
functions. It should be emphasized though that $X_L$ and $p_L$ are not
operators and can be treated as ordinary classical objects.

We point out that there is no need to invoke any {\em ad hoc}
thermodynamic analogy (e.g., fluctuation-dissipation relations) in our
derivation of the stochastic quantum Liouville equation as was done by
Graziani \cite{kn:graz} in a first attempt to apply the Wigner
function formalism to stochastic inflation. As we will show in the
following sections, such relations do not hold in general and any
analogy with conventional Brownian motion must be treated with extreme
caution. A related remark is that since (\ref{asixty}) is formally a
master equation one might expect to define a suitable entropy
satisfying some variant of the H-theorem \cite{kn:shhk}. For example,
it is easy to see that because of the diffusion term, the ``linear
entropy'' or ``mixing parameter'' $Tr\r^2=\int dX_Ldp_L~f^{~2}_W$ will
always decrease with time (implying that the quantum state is getting
more and more mixed). This must not be interpreted in the sense of
``quantum decoherence'' \cite{kn:deco} as we are dealing with a {\em
free} theory and there is no coupling to some external environment.
(One way to understand this result may be that this decrease simply
mirrors the loss of information inherent in our time dependent
coarse-graining.)

Finally we draw attention to some technical issues. Note that no
assumption is needed as to the symmetry properties of $B_{ij}(\h)$;
this allows for the fact that the noises do not commute. Note also
that while separately $B_{12}$ and $B_{21}$ need not be real, they
appear in (\ref{asixtya}) only in the symmetrized combination
$B_{12}+B_{21}$ (since partial derivatives commute), which, as is
clear from (\ref{athirtysix}) and (\ref{athirtyseven}) is always real.
The derivation of (\ref{asixty}) is also free from any kind of
``slow-roll'' assumption although this merit is mainly technical as
physical results at late times remain unaffected when such conditions
are imposed (see Ref. \cite{kn:tgp}, Appendix A).

We now turn to the problem of solving the stochastic quantum Liouville
equation. Formally, the solutions are not difficult to obtain as
(\ref{asixty}) is just a Kramers equation describing a time dependent
Ornstein-Uhlenbeck process \cite{kn:ngvk}. The average values satisfy
\bq
{d\over d\h}\VEV{z_i}_N=A_{ij}(\h)\VEV{z_j}_N  \label{asixtyone}
\eq
given the initial condition $\VEV{z_i(\h_0)}_N=z_{i0}$. The matrix
$A_{ij}$ is defined by
\bq
L_0f_W=A_{ij}{\pa\over\pa z_i}\left(z_jf_W\right) \label{asixtytwo}
\eq
and in our case, $A_{11}=A_{22}=0,~A_{12}=1,~A_{21}=-\o^2(\h)$. The
propagator for the average values $\VEV{z_i}$, $G_{ij}$, satisfies,
\bq
{d\over d\h}G_{ij}=A_{ik}G_{kj};~~~~~~~G_{ij}(\h_0)=\d_{ij}.
\label{asixtytwob}
\eq
The second moments follow from
\bq
{d\over d\h}\VEV{z_iz_j}_N=A_{ik}\VEV{z_kz_j}_N+
A_{jk}\VEV{z_iz_k}_N+D_{ij}.        \label{asixtythree}
\eq
This equation is also obeyed by the covariances
$\Xi_{ij}=\VEV{z_iz_j}_N-\VEV{z_i}_N\VEV{z_j}_N$, which can themselves
be expressed in terms of the propagator as
\bq
\Xi(\h)=G(\h)\cdot\Xi(\h_0)\cdot\tilde{G}(\h) + \int_{\h_0}^{\h}d\h'~G(\h)\cdot
G^{-1}(\h')\cdot D(\h')\cdot\tilde{G}^{-1}(\h')\cdot\tilde{G}(\h).
\label{asixtythreeb}
\eq
The first term is the systematic contribution arising from a
reversible dynamical evolution from the given initial condition. The
second term represents the irreversible stochastic contribution due to
the diffusion matrix.

It can be shown \cite{kn:ngvk} that the general solution of
(\ref{asixty}) with delta function initial conditions
$W(z,\h_0)=\prod_i\d(z_i-z_{i0})$ is
\bq
W(z,\h;z_0,\h_0)={1\over 2\p}(Det~\Xi)^{-1/2}\exp\left(-{1\over
2}(\tilde{z}-\VEV{\tilde{z}}_N)\cdot\Xi^{-1}\cdot
(z-\VEV{z}_N)\right).                     \label{asixtyfour}
\eq
The function $W(z,\h;z_0,\h_0)$ serves as the propagator for the
Fokker-Planck equation (\ref{asixty}). Solutions for arbitrary initial
conditions $f_W(z_0,\h_0)$ can be generated from it by
\bq
f_W(z,\h)=\int dz_0~W(z,\h;z_0,\h_0)f_W(z_0,\h_0).   \label{asixtyfive}
\eq
These general results will be applied to exponential and  power law
expansions in the following sections.
\newpage

\centerline{\bf III. Exponential Inflation}

The case of a de Sitter expansion furnishes a particularly simple
example in which to implement the procedures of Sec. II. Our stochastic
approach not only reproduces some previous field theoretic results but
also introduces a new interpretive framework. In this section we aim
mainly to obtain quantitative results.

The scale factor for de Sitter space is
\bq
S(\h)=-{1\over H_0\h},                            \label{cone}
\eq
and the conformal time
\bq
\h=-{1\over H_0}\hbox{e}^{-H_0t}.                  \label{ctwo}
\eq
We note that here we are not really interested in the case of an
eternal de Sitter expansion. Initial conditions for stochastic
inflation will be assigned in the finite past, at the beginning of the
inflationary phase.

The scalar field modes now satisfy
\bq
\ddot{\c}_k + \left(k^2 + {1\over\h^2}\left({m^2\over
H_0^{~2}}-2\right)\right)\c_k=0.                 \label{cthree}
\eq
If the mass is zero or at least small compared to $H_0^{~2}$, the
``unstable'' sector is characterized by $k^2 < 2/\h^2$. Therefore,
following Starobinsky \cite{kn:aas} it makes sense to set
\bq
k_S(\h)={\e\over\h}                          \label{cthreea}
\eq
where $\e$ is a constant that serves to parametrize the cutoff. If we
assume that inflation began at the time $\h_0$, this implying a
natural infrared cutoff $\h_0^{-1}$ (more details may be found in
the Appendix), it is clear that $\e$ cannot be arbitrarily small as we
must have $\e\h^{-1}>\h_0^{-1}$. If one is interested only in late
time results, i.e., when $\h\ll\h_0$ then $\e$ may be taken to be
small. In this paper we will not restrict $\e$ to be arbitrarily small
but will allow it to be as large as unity. In principle, it is
desirable that physical answers not depend on $\e$; this will turn out
{\em not} to be the case. As will be shown later all infrared
divergent quantities are only weakly dependent on $\e$ but this does
not hold in general (the situation is more complicated for power law
inflation). Our stochastic approach will correctly reproduce the cutoff
dependences for infrared finite quantities calculated from
conventional quantum field theory (see the Appendix).

A curious special property of de Sitter space is that even when
the mass is non-zero (no infrared divergence), there is
still an initial growth of $\VEV{\F^2}$ to an asymptotic limit
$\VEV{\F^2}_{BD}$, which is the value in the Bunch-Davies vacuum
\cite{kn:destab}. The mathematical reason for this behavior is simply
that the mass and curvature contributions in (\ref{cthree}) scale
identically with conformal time and that for $m^2$ small compared to
$H_0^{~2}$, there is still an ``unstable'' infrared regime despite
there being no infrared divergence. As will be made clear in the next
section this feature is not shared by power law inflation.

The mode equation (\ref{cthree}) admits the general solution
\bq
\c_k(\h)=C_1\h^{1/2}\hbox{e}^{i\n\p/2}H_{\n}^{(1)}(k\h)+
C_2\h^{1/2}\hbox{e}^{-i\n\p/2}H_{\n}^{(2)}(k\h),   \label{cfour}
\eq
where
\bq
\n^2={9\over 4}-{m^2\over H_0^{~2}}.                   \label{cfive}
\eq
The arbitrariness of the de Sitter vacuum is reflected in the various
possible choices for $C_1$ and $C_2$. In this paper the quantum state
we will use is the Bunch-Davies vacuum \cite{kn:bad}, characterized by
$C_1=0$ and $C_2=\sqrt{\p}/2$.

The symmetrized diffusion matrix $D_{ij}$ now follows from
(\ref{athirtyfive})--(\ref{athirtyeight}):
\ba
D_{11}(\h_1)=2B_{11}(\h_1)&\simeq&{\e^3\over 8\p}\h_1^{-3}
\abs{H_{\n}^{(2)}(\e)}^2,                          \label{csix}\\
D_{12}(\h_1)=B_{12}(\h_1)+B_{21}(\h_2)&\simeq&{\e^3\over 8\p}
\h_1^{-4}Re\left\{H_{\n}^{(2)}(\e)\left[\left({1\over
2}-\n\right){H_{\n}^{(2)}}^*(\e)
+\e{H_{\n-1}^{(2)}}^*(\e)\right]\right\}          \nonumber\\
&\simeq& D_{21}(\h_1),                         \label{ceight}  \\
D_{22}(\h_1)=2B_{22}(\h_1)&\simeq&{\e^3\over
8\p}\h_1^{-5}\abs{\left({1\over
2}-\n\right)H_{\n}^{(2)}(\e)+\e H_{\n-1}^{(2)}(\e)}^2. \label{cnine}
\ea

Our eventual goal is to solve the Fokker-Planck equation
(\ref{asixty}) given the diffusion coefficients
(\ref{csix})--(\ref{cnine}). The potential term in the systematic
component (\ref{afortynine}) of the stochastic Liouville operator is
characterized by
\bq
\o^2(\h)=\left({1\over 4}-\n^2\right){1\over \h^2}.  \label{cten}
\eq
For all the cases we consider in this paper $\o^2(\h)$ will be
negative (as $\n^2>1/4$). We are dealing therefore with a time
dependent upside down harmonic oscillator. The equation for the
propagator ({\ref{asixtytwob}) can now be easily solved, and in terms
of the initial time $\h_0$, we find
\ba
G_{11}(\h)&=&{1\over 2\n}\left[(\n-1/2)\left({\h\over
\h_0}\right)^{\n+1/2} + (\n+1/2) \left({\h\over
\h_0}\right)^{-\n+1/2}\right],        \label{celeven}\\
G_{12}(\h)&=&{(\h\h_0)^{1/2}\over 2\n}\left[\left({\h\over
\h_0}\right)^{\n}-\left({\h\over \h_0}\right)^{-\n}\right],
\label{ctwelve}\\
G_{21}(\h)&=&{(\n^2-1/4)\over 2\n}(\h\h_0)^{-1/2}\left[\left({\h\over
\h_0}\right)^{\n}-\left({\h\over \h_0}\right)^{-\n}\right],
\label{cthirteen}\\
G_{22}(\h)&=&{1\over 2\n}\left[(\n+1/2)\left({\h\over
\h_0}\right)^{\n-1/2}+(\n-1/2)\left({\h\over
\h_0}\right)^{-\n-1/2}\right].        \label{cfourteen}
\ea
This is the general solution, valid for all the special cases
considered in this paper.

We now confine attention to the massless case where the parameter
$\n=3/2$, and
\ba
H_{1/2}^{(2)}(k\h)&=&i\left(2k\h\over\p\right)^{1/2}{\hbox{e}^{-ik\h}\over
k\h},                                              \label{cfifteen}\\
H_{3/2}^{(2)}(k\h)&=&\left(2k\h\over\p\right)^{1/2}{\hbox{e}^{-ik\h}\over
(k\h)^2}(i-k\h).                               \label{csixteen}
\ea
The diffusion matrix then takes the simple form,
\ba
D_{11}(\h_1)&\simeq&{1\over 4\p^2}\h_1^{-3}\left(1+\e^2\right),
\label{cseventeen}\\
D_{12}(\h_1)=D_{21}(\h_1)&\simeq&-{1\over 4\p^2}\h_1^{-4},
\label{ceighteen}\\
D_{22}(\h_1)&\simeq&{1\over 4\p^2}\h_1^{-5}\left(1-\e^2+\e^4\right).
\label{cnineteen}
\ea
Note that to leading order (for $\e\ll 1$) the diffusion matrix is
independent of $\e$. It is misleading however to conclude that the
actual value of $\e$ is not important as at this order
$Det~\Xi_{ij}=0$ (notice that this is a direct consequence of
retaining noise cross-correlations). In order to eventually obtain a
nonsingular covariance matrix we must go beyond this level of
approximation; the final solution for the Wigner function is in fact
strongly dependent on $\e$.

It is a tedious but straightforward exercise to obtain the covariance
matrix using (\ref{asixtythreeb}). We assume an initial distribution
such that $z_{i0}=0$ but impose no conditions on $\Xi_{ij}(\h_0)$.
Ignoring for the moment the systematic component, the stochastic
contribution turns out to be
\ba
\Xi_{11}&=&{\ln{(\h_0/\h)}\over 4\p^2\h^2}\left(1+{\e^2\over
3}+{\e^4\over 9}\right)+{\e^2\over 18\p^2\h^2}\left(1-{\e^2\over
4}\right)-{\e^2\h\over 18 \p^2\h_0^{~3}}\left(1-{\e^2\over
3}+{\h^3\over 12\h_0^{~3}}\right),\nonumber \\&& \label{ctwenty}\\
\Xi_{12}&=&{\ln{(\h_0/\h)}\over
4\p^2\h^3}\left(1+{\e^2\over3}+{\e^4\over 9}\right)+{\e^2\over
36\p^2\h^3}-{\e^2\over 36\p^2\h_0^{~3}}\left(1-{\e^2\over
3}+{\e^2\h^3 \over 3\h_0^{~3}}\right)\nonumber\\
&=&\Xi_{21},\label{ctwentyone}\\
\Xi_{22}&=&{\ln{(\h_0/\h)}\over
4\p^2\h^4}\left(1+{\e^2\over3}+{\e^4\over 9}\right)-{\e^2\over
9\p^2\h^4}\left(1-{\e^2\over 2}\right)+{\e^2\over
9\p^2\h_0^{~3}\h}\left(1-{\e^2\over 3}-{\e^2\h^3\over
6\h_0^{~3}}\right).     \nonumber\\       \label{ctwentytwo}
\ea
One can easily check using (\ref{asixtythreeb}) and
(\ref{celeven})--(\ref{cfourteen}) that at late times the systematic
contribution to the covariance matrix is negligible compared to the
stochastic piece. (Initial conditions are discussed further below.)

The full solution for the noise averaged Wigner distribution function
follows trivially from (\ref{asixtyfour}) and (\ref{asixtyfive}). In
this section we will concentrate only on the covariance matrix itself
as all average values of interest can be computed directly from it.
Detailed study of the distribution function will be postponed to Sec.
V.

The covariance matrix (\ref{ctwenty})--(\ref{ctwentytwo}) refers to
the ``conformal'' variables $X_L$ and $p_L$. Reverting to the original
field $\F$, we introduce new c-number variables $\f_c$ and $p_c$ via
the canonical transformation
\ba
\f_c&=&{X_L\over S},          \label{ctwentythree}\\
p_c&=&S p_L-\dot{S} X_L.  \label{ctwentythreea}
\ea
The corresponding covariance matrix may be written as
\ba
\Xi^{(\f)}_{11}=\VEV{\f_c^{~2}}_N&=&S^{-2}\Xi_{11}, \label{ctwentyfour}\\
\Xi^{(\f)}_{12}=\Xi^{(\f)}_{21}=\VEV{\f_c p_c}_N&=&\Xi_{12}-{\dot{S}\over
S}\Xi_{11}, \label{ctwentyfive}\\
\Xi^{(\f)}_{22}=\VEV{p_c^{~2}}_N&=&S^2 \Xi_{22}-2\dot{S}
S\Xi_{12}+\dot{S}^2 \Xi_{11}, \label{ctwentysix}
\ea
where all averages of the type $\VEV{z_i(\h)}$ vanish as a consequence
of our choice $z_{i0}=0$ for the initial condition. Comparison with
the field theoretic results is simpler if we introduce the
``velocity'' $\f_c'$ (the prime denotes differentiation with respect
to cosmic time) in place of the canonical momentum $p_c$. Noting that for
de Sitter space, $\ln{\h_0/\h}=H_0(t-t_0)$, where $t_0$ denotes the
beginning of the exponential expansion, and using
(\ref{ctwenty})--(\ref{ctwentytwo}), the new covariance matrix (for a
massless field) turns out to be
\ba
\VEV{\f_c^{~2}}_N&=&\Xi^{(\f)}_{11}= {H_0^{~3}\over
4\p^2}(t-t_0)\left(1+{\e^2\over
3}+{\e^4\over 9}\right)+{\e^2H_0^{~2}\over 18\p^2}\left(1-{\e^2\over
4}\right)\nonumber\\
&&-{\e^2H_0^{~2}\over
18\p^2}\hbox{e}^{-3H_0(t-t_0)}\left(1-{\e^2\over 3}+{\e^2\over
12}\hbox{e}^{-3H_0(t-t_0)}\right), \label{ctwentyseven}\\
\VEV{\f_c\f_c'}_N&=&{\Xi^{(\f)}_{12}\over
S^3}=\Xi^{(\f)}_{21}={\e^2H_0^{~3}\over 12\p^2} \left(1-{\e^2\over
6}\right)\nonumber\\
&& - {\e^2H_0^{~3}\over 12\p^2}\hbox{e}^{-3H_0(t-t_0)}
\left(1-{\e^2\over 3}+{\e^2\over 6}\hbox{e}^{-3H_0(t-t_0)}\right),
\label{ctwentyeight}\\
\VEV{\f_c'^2}_N&=&{\Xi^{(\f)}_{22}\over S^6}={\e^4H_0^{~4}\over
24\p^2}\left(1-\hbox{e}^{-6H_0(t-t_0)}\right).
\label{ctwentynine}
\ea
The above results record only the stochastic contribution to the
covariance matrix. It is easy to compute the systematic contribution
for an arbitrary initial choice of $\Xi_{ij}$ from
(\ref{asixtythreeb}) and (\ref{celeven})--(\ref{cfourteen}) (since the
Wigner distribution function must be square integrable we cannot take the
initial distribution to be a delta function over phase space). The
contribution to $\Xi_{11}$ consists of a constant piece and terms that
fall off exponentially with cosmic time. Contributions to $\Xi_{12}$
and $\Xi_{22}$ also display a similar exponential fall-off. It is
important to note that these contributions, though insignificant at
late times, {\em can} dominate similar terms that already exist in the
stochastic piece (especially for small values of $\e$). Therefore,
{\em only} the late time limit is independent of initial conditions.
This is in contrast to the field theoretic case where the contribution
from initial conditions is usually irrelevant even at early times (see
the Appendix). We also draw attention to the fact that the exponential
fall-offs in the stochastic calculation are not the same as the field
theoretic ones; again, this is of no consequence at late times.

As long as the initial distribution is such that $\VEV{\f_c}_N$ and
$\VEV{p_c}_N$ are zero (i.e., $z_{i0}=0$), at late times ($\h$ small)
and with $\e\ll 1$, the leading order contributions are
\ba
\Xi^{(\f)}_{11}&=& \VEV{\f_c^{~2}}_N\simeq {H_0^{~3}\over
4\p^2}(t-t_0),                         \label{cthirty}\\
{\Xi^{(\f)}_{12}\over S^3}&=& \VEV{\f_c \f_c'}_N\simeq {\e^2H_0^{~3}\over
12\p^2},                          \label{cthirtyone}\\
{\Xi^{(\f)}_{22}\over S^6}&=& \VEV{\f_c'^{~2}}_N\simeq {\e^4H_0^{~4}\over
24\p^2}.                          \label{cthirtytwo}
\ea
The $\Xi_{11}^{(\f)}$ term reproduces the standard quantum field
theoretic result \cite{kn:destab} (and (\ref{Afive}), the Appendix)
for the expectation value $\VEV{\F^2}$, here viewed as a noise average
for the c-number variable $\f_c$ {\em provided} that $\e$ is small.
However, if we set $\e\sim 1$ the answer does not agree with the field
theoretic result (\ref{Afive}) found in the Appendix (which unlike the
stochastic calculation is essentially cutoff independent).

For the sake of comparison, if we set $\e=1$ in
(\ref{ctwentyseven})--(\ref{ctwentynine}) we find, at late times,
\ba
\Xi^{(\f)}_{11}&=& \VEV{\f_c^{~2}}_N\simeq {13 H_0^{~3}\over 36
\p^2}(t-t_0),                          \label{cthirtyb}\\
{\Xi^{(\f)}_{12}\over S^3}&=& \VEV{\f_c \f_c'}_N\simeq {5
H_0^{~3}\over 72\p^2},
\label{cthirtyoneb} \\
{\Xi^{(\f)}_{22}\over S^6}&=& \VEV{\f_c'^{~2}}_N\simeq {H_0^{~4}\over
24\p^2},               \label{cthirtytwob}
\ea
whereas the corresponding field theoretic results (\ref{Afive}),
(\ref{Anine}), and (\ref{Atwelve}) of the Appendix give,
at late times:
\ba
\VEV{\F^2}&\simeq& {H_0^{~3}\over 4\p^2}(t-t_0), \label{cthirtyc}\\
{1\over 2}\VEV{\F\F'+\F'\F}&\simeq& {\e^2H_0^{~3}\over 8\p^2},
\label{cthirtyonec}\\
\VEV{\F'^2}&\simeq& {\e^4H_0^{~4}\over16 \p^2}.  \label{cthirtytwoc}
\ea
We see that while the stochastic results for $\Xi^{(\f)}_{12}$
and $\Xi^{(\f)}_{22}$ correctly reproduce the the cutoff dependence
found in the field theoretic case, the numerical values of the
coefficients do not match. This is not a serious problem as these
quantities need to be renormalized anyway (something that is beyond
the scope of this paper).

The technical reason for the disagreement between the field theoretic
and stochastic calculations is the neglect of spatial derivatives in
(\ref{athirtyfour}). One is attempting to approximate a time dependent
quantum sum over modes by a modified dynamics (via the noise term) for
the zero mode and neglecting all the other modes (apart from their
contribution to the noise). When computing $\VEV{\F^2}$ via the
standard field theoretic method the infrared sector provides the
dominant contribution. On the other hand, when computing
$\VEV{\F\F'+\F'\F}/2$ and $\VEV{\F'^2}$ extra multiplicative factors of
$k$ and $k^2$ (see the Appendix) weaken this infrared dependence. One
expects therefore that the stochastic method should work better for
$\VEV{\F^2}$ and, as we have seen, this is indeed the case. The fact
that stochastic results are more accurate for small values of $\e$ is
also easy to appreciate: a small $\e$ means that only long wavelength
modes are contributing to the noise so that $k$ is indeed small and
can be neglected. However, if we let $\e$ be of order unity, then the
neglect of spatial derivatives will lead to errors. A possible remedy
is to work with a Wigner functional defined directly from the field
theory but this may well be at the expense of the calculational ease
that characterizes the present approach.

We now consider the massive field but confine attention to the case
$m^2\ll H_0^{~2}$. The propagator $G(\h)$ is still given by
(\ref{celeven})--(\ref{cfourteen}) except that the parameter $\n$ is
now given by
\bq
\n\simeq{3\over 2}-{m^2\over 3H_0^{~2}}.   \label{cthirtythree}
\eq
In the limit of a small mass the diffusion matrix is essentially the
same as for the massless case (since $\n\sim 3/2$). Keeping the
diffusion matrix given by (\ref{cseventeen})--(\ref{cnineteen}) but
using (\ref{cthirtythree}) for the propagator it is a simple matter to
solve for the stochastic contribution to the covariance matrix. The
final expressions are very long and not very illuminating. Here we
present only the leading order stochastic terms at late times:
\ba
\VEV{\f_c^{~2}}_N&\simeq& {3H_0^{~4}\over
8m^2\p^2}\left[1-\exp{\left(-{2m^2\over
3H_0}(t-t_0)\right)}\right]\left[1+{\e^2\over 3}+{\e^4\over 9} +
{m^2\over 9H_0^{~2}}\right],
\label{cthirtyfour}\\
\VEV{\f_c \f_c'}_N&\simeq& {H_0^{~3}\over
8\p^2}\left[1+\e^2-\left(1+{\e^2\over 3}+{\e^4\over
9}\right)\exp{\left(-{2m^2\over 3H_0}(t-t_0)\right)}\right],
\label{cthirtyfive}\\
\VEV{\f_c'^{~2}}_N&\simeq& {H_0^{~4}\over
24\p^2}\left[\e^4+{m^2\over H_0^{~2}}\left(1-\exp{\left(-{2m^2\over
3H_0}(t-t_0)\right)}\right)\right.\nonumber\\
&&\left.+{\e^2m^2\over H_0^{~2}}\left(1-\exp{\left(-{2m^2\over
3H_0}(t-t_0)\right)}\right)\right].    \label{cthirtysix}
\ea
where we have dropped all terms that vanish faster at late times
and also neglected terms that are of higher order in $m^2/H_0^{~2}$.
An interesting feature of the massive case is that all contributions
arising from $\Xi_{ij}^{(\f)}(\h_0)$ are negligible even at early
times (unlike the massless case). The agreement with field theory is
remarkably good. With $\e$ small, or more precisely, in the range,
\bq
\exp(-3H_0^{~2}/(2m^2))\ll\e^2\ll m^2/H_0^{~2}, \label{masscon}
\eq
all asymptotic late time values are exactly reproduced (similar
inequalities are derived somewhat differently in Refs. \cite{kn:aas}
and \cite{kn:sas}) . Unlike the massless case, this time the approach
to these late time values is also in agreement with the field
theoretic results (i.e., no mismatch in the exponentially falling off
terms). At late times the above expressions reduce to
\ba
\Xi_{11}^{(\f)}=\VEV{\f_c^{~2}}_N&\simeq& {3H_0^{~4}\over
8m^2\p^2}\left[1+{m^2\over 9H_0^{~2}}-\exp{\left(-{2m^2\over
3H_0}(t-t_0)\right)}\right],
\label{cthirtyseven}\\
{\Xi_{12}^{(\f)}\over S^3}=\VEV{\f_c \f_c'}_N&\simeq& {H_0^{~3}\over
8\p^2}\left[1-\exp{\left(-{2m^2\over 3H_0}(t-t_0)\right)}\right],
\label{cthirtyeight}\\
{\Xi_{22}^{(\f)}\over S^6}=\VEV{\f_c'^{~2}}_N&\simeq& {m^2H_0^{~2}\over
24\p^2}\left[1-\exp{\left(-{2m^2\over 3H_0}(t-t_0)\right)}\right],
\label{cthirtynine}
\ea
in agreement with the field theoretic results (\ref{Aseventeen}),
(\ref{Atwentyone}), and (\ref{Atwentythree}) of the Appendix (with $\e$
in the previously indicated range). Unlike the massless case there is
no leading order $\e$ dependence in $\Xi_{12}$ and $\Xi_{22}$. To
avoid $Det~\Xi_{ij}^{(\f)}=0$, it is important to keep the
subdominant mass squared term in (\ref{cthirtyseven}). It is again
easy to verify that agreement with field theoretic results does not
extend to the case $\e\sim 1$; the two calculations now differ by
multiplicative factors of order unity.

\newpage

\centerline{\bf IV. Power Law Inflation}

In this section we treat a power law expansion $a(t)\sim t^p$ with
$p>1$ and consider only the case of a massless scalar field. Assuming
that inflation set in at the initial time $\h_0$ (with the scale
factor set to unity at this time), we find from (\ref{athree}),
\bq
S(\h)=\left({\h\over\h_0}\right)^{(1-2\n)/2}     \label{bone}
\eq
where
\bq
\n={1-3p\over 2(1-p)}.                       \label{btwo}
\eq
It then follows that
\bq
{\ddot{S}\over S}=-\left({1\over 4}-\n^2\right){1\over\h^2}.  \label{bthree}
\eq
It is useful to note that for $p>1$, $\n>3/2$ and also that as
$t\rightarrow \infty$, $\h\rightarrow 0$. In the formal limit
$p\rightarrow \infty$, $\n=3/2$ which is the value for de Sitter
space.

The scalar field modes now satisfy
\bq
\ddot{\c}_k+\left(k^2+ m^2\left(\h\over\h_0\right)^{1-2\n}+\left({1\over
4}-\n^2\right){1\over\h^2}\right)\c_k=0.       \label{bfour}
\eq
This time the mass and curvature contributions scale differently with
conformal time. It is easy to see that at late times the mass term
dominates the curvature contribution since $\n>3/2$. Therefore, as
cosmic time increases (and the conformal time decreases) there is a
continuous flow {\em from} the ``unstable'' to the ``stable'' sector.
However, destabilization will still occur for the massless case where
the ``unstable'' sector is characterized by $k^2 < (\n^2-1/4)/\h^2$.
We see also that even in the case of power law inflation it is only
natural to implement the same choice that we made in the last section,
i.e., to set $k_S(\h)={\e/\h}$.

The mode equation (\ref{bfour}) for a massless field admits the
general solution
\bq
\c_k(\h)=C_1\h^{1/2}H^{(1)}_{\n}(k\h)+C_2\h^{1/2}H^{(2)}_{\n}(k\h)
\label{bfive}
\eq
or, in terms of the original field,
\bq
\f_k(\h)=C_1\h_0^{~1/2}\left(\h\over\h_0\right)^{\n}H^{(1)}_{\n}(k\h)+
C_2\h_0^{~1/2}\left(\h\over\h_0\right)^{\n}H^{(2)}_{\n}(k\h).
\label{bsix}
\eq
The adiabatic vacuum is specified by $C_1=0,~C_2=\sqrt{\p}/2$, i.e.,
\bq
\c_k(\h)=\left({\p\h\over 4}\right)^{1/2}H^{(2)}_{\n}(k\h).    \label{bseven}
\eq
Assuming that all ``high frequency'' modes were in the adiabatic
vacuum at the onset of inflation, (\ref{bseven}) enables us to compute
the diffusion coefficients (\ref{athirtyfive})--(\ref{athirtyeight})
which are the same as (\ref{csix})--(\ref{cnine}) except that now $\n$
is specified by (\ref{btwo}).

Power law inflation with $p\gg 1$ can be treated in a simple and
direct manner by following the same approach as that for the massive
field in exponential inflation. Note that when $p$ is large,
\bq
\n\simeq {3\over 2} + {1\over p} + {1\over p^2} + {1\over p^3} +
\cdots.         \label{bpone}
\eq
This allows us to approximate the diffusion matrix by the one for
$\n=3/2$, (\ref{cseventeen})--(\ref{cnineteen}). The propagator
$G(\h)$ is given by (\ref{celeven})--(\ref{cfourteen}) with $\n$
specified by (\ref{bpone}). The stochastic piece of the covariance
matrix can now be found by a straightforward computation. The result
is too long to write out in entirety and we content ourselves by just
displaying the leading order terms:
\ba
\Xi_{11}^{(\f)}=\VEV{\f_c^{~2}}_N&\simeq&{p\over
8\p^2}\h_0^{-2}\left[1-\left({\h\over
\h_0}\right)^{2/p}\right]\left[1+{1\over 3}\e^2+{1\over
9}\e^4\right]\\         \label{bptwo}
&\simeq&{p\over 8\p^2}\h_0^{-2}\left[1-\left({t_0\over
t}\right)^2\right]\left[1+{1\over 3}\e^2+{1\over 9}\e^4\right],
\label{bpthree}\\
{\Xi_{12}^{(\f)}\over S^3}=\VEV{\f_c \f_c'}_N&\simeq&{1\over
12\p^2}\h_0^{-3}\left({\h\over\h_0}\right)^{3/p}\left(\e^2 + {1\over
p}\right)                                    \label{bpfour}\\
&\simeq&{1\over 12\p^2}\h_0^{-3}\left({t_0\over t}\right)^3\left(\e^2
+ {1\over p}\right),                                    \label{bpfive}\\
{\Xi_{22}\over S^6}=\VEV{\f_c'^{~2}}&\simeq&{1\over
12\p^2}\h_0^{-4}\left({\h\over\h_0}\right)^{4/p}\left[{1\over 2p^2} +
\e^2\left({1\over p} + {\e^2\over 2}\right)\right]      \label{bpsix}\\
&\simeq&{1\over 12\p^2}\h_0^{-4}\left({t_0\over
t}\right)^{4}\left[{1\over 2p^2} + \e^2\left({1\over p} + {\e^2\over
2}\right)\right],      \label{bpseven}
\ea
where terms lower order in $1/p$ and vanishing faster at late times
have been dropped. The variables $\f_c$ and $p_c$ are still defined
by (\ref{ctwentythree}) and (\ref{ctwentythreea}) except that $S(\h)$
is now given by (\ref{bone}). For small $\e$, (\ref{bpthree}) is in
agreement with the field theoretic calculation of \cite{kn:despl}
(also compare with (\ref{Athirtyone}) of the Appendix). However, just
as for the massive field in de Sitter space, $\e$ {\em cannot} be
arbitrarily small. Consistency with the field theoretic results
(\ref{Athirtyeight}) and (\ref{Afortytwo}) of the Appendix requires
that
\bq
p^{-1}\ll\e^2\ll 1
\eq
in (\ref{bpfive}) and (\ref{bpseven}). The above expectation values
were also computed by Kandrup using a different method \cite{kn:pli2}.
While our result for $\VEV{\f_c^{~2}}$ is in agreement with his, this
is not true for the other two cases. The inconsistency can be traced
to an approximation for the noise that does not take the commutator
properly into account (see the discussion of this point in Sec. II).

In this case, while at late times $\VEV{\f_c^{~2}}_N$ goes to a
constant (as for the massive field in de Sitter space),
$\VEV{\f_c\f_c'}_N$ and $\VEV{\f_c'^{~2}}_N$ vanish. This is in
contrast with the case of a massive field in de Sitter space where
these quantities instead of vanishing, also go to constant values.
The role of initial conditions is similar to that for the massive
field in de Sitter space rather than the massless one: the systematic
contribution to the covariance matrix is always negligible as long as
$p\gg 1$ (but not otherwise). Consequently, the late time results
follow from (\ref{bptwo})--(\ref{bpseven}):
\ba
\VEV{\f_c^{~2}}_N&\simeq&{p\over 8\p^2}\h_0^{-2},  \label{blt}\\
\VEV{\f_c \f_c'}_N&\simeq&{\e^2\over 12\p^2}\h_0^{-3}\left({t_0\over
t}\right)^3,                            \label{bltb}\\
\VEV{\f_c'^{~2}}_N&\simeq&{\e^4\over 24\p^2}\h_0^{-4}\left({t_0\over
t}\right)^4,                          \label{bltc}
\ea
where we have taken $\e\ll 1$. For $\e=1$, we have,
\ba
\VEV{\f_c^{~2}}_N&\simeq&{13 p\over 72\p^2}\h_0^{-2},  \label{blta}\\
\VEV{\f_c \f_c'}_N&\simeq&{5\over 72\p^2}\h_0^{-3}\left({t_0\over
t}\right)^3,                            \label{bltab}\\
\VEV{\f_c'^{~2}}_N&\simeq&{1\over 24\p^2}\h_0^{-4}\left({t_0\over
t}\right)^4.                          \label{bltac}
\ea

The field theoretic results (\ref{Athirtyone}), (\ref{Athirtynine}),
and (\ref{Afortythree}) of the Appendix yield the corresponding late
time limits,
\ba
\VEV{\F^2}&\simeq&{p\over 8\p^2}\h_0^{-2},  \label{bqlt}\\
{1\over 2}\VEV{\F\F'+\F'\F}&\simeq&{\e^2\over 8\p^2}\h_0^{-3}\left({t_0\over
t}\right)^3,                            \label{bqltb}\\
\VEV{\F'^2}&\simeq&{\e^4\over 16\p^2}\h_0^{-4}\left({t_0\over
t}\right)^4.                          \label{bqltc}
\ea
There is reasonable agreement with the stochastic results when $\e$ is
small but as is expected the results diverge from each other when
$\e\sim 1$.

The limit $p\rightarrow\infty$ may be applied to (\ref{bptwo}) using
\bq
\left({\h\over\h_0}\right)^{2/p}=1+{2\over
p}\ln\left({\h\over\h_0}\right)+\cdots,   \label{bpeight}
\eq
with the result
\bq
\VEV{\f_c^{~2}}_N\simeq {H_0(t-t_0)\over
4\p^2\h_0^{~2}}\left[1+{1\over 3}\e^2+{1\over 9}\e^4\right]. \label{bpnine}
\eq
Since $H_0=\h_0^{-1}$, this agrees with the result
(\ref{ctwentyseven}) for a massless field in de Sitter space. In a
similar manner one can check that $\VEV{\f_c \f_c'}_N$ and
$\VEV{\f_c'^{~2}}_N$ also reduce to the appropriate expressions for a
massless field in de Sitter space as calculated from the stochastic
approach.
\newpage

\centerline{\bf V. Solutions and Interpretations}

In this section we study the full phase space distribution function.
Given that we have already computed the relevant covariance matrices
it is now a simple matter to write out the corresponding Wigner
functions. In the examples studied here these distributions will be
positive definite and as such may be interpreted as true probability
distributions, at least formally.

A knowledge of the distribution function is important as it will
enable us to critically address issues such as the existence of
fluctuation-dissipation relations and whether there exist late time
thermal solutions or not. These are the problems we will tackle first.

The stochastic Liouville equation (\ref{asixty}) is written in terms
of the conformal variables $X_L$ and $p_L$. In all the examples we
studied, $z_{i0} = 0$, in which case the general solution
(\ref{asixtyfour}) becomes, at late times,
\bq
f_W(z,\h) = {1\over 2\p}(Det~\Xi)^{-1/2}\exp{\left(-{1\over
2}\tilde{z}\cdot\Xi^{-1}\cdot z\right)},        \label{csone}
\eq
all contributions from nontrivial initial conditions having washed out
in this limit. Converting to the variables $\f_c$ and $p_c$
appropriate to the original frame, the distribution function
$(\ref{csone})$ goes over to
\bq
f_{cl}(\f_c,p_c) = {1\over
2\p}(Det~\Xi^{(\f)})^{-1/2}\exp{\left(-{1\over
2}\left[{\Xi_{11}^{(\f)}}^{-1}\f_c^{~2} + 2{\Xi_{12}^{(\f)}}^{-1}\f_c
p_c +{\Xi_{22}^{(\f)}}^{-1}p_c^{~2}\right]\right)}.
\label{cstwo}
\eq
It is important to appreciate that while $f_{cl}$ gives the
correct expectation values (Secs. III and IV) and is a perfectly
respectable classical distribution, it is not a Wigner function
defined from the beginning for the variables $\f_c$ and $\p_c$. This
is because, as we noted earlier, these distributions are not invariant
under canonical transformations. (We are treating the conformal
variables $\c_L$ and $\p_L$ as the preferred variables to quantize.)
However, the key point is that in our case the linear entropy remains
invariant under this transformation.

A key observation regarding (\ref{cstwo}) is that a knowledge of the
reduced distribution
\ba
f_{r}(\f_c)&\equiv&\int_{-\infty}^{+\infty}dp_c f_W(\f_c,p_c)
\nonumber\\
&=&{1\over\sqrt{2\p}}\left[\Xi_{11}^{(\f)}\right]^{-1/2}\exp\left(-{1\over
2}{\f_c^{~2}\over \Xi_{11}^{(\f)}}\right) \label{cstwored}
\ea
is of no use in reconstructing the original distribution. This trivial
fact has important consequences if one attempts thermodynamic
interpretations of the results from our stochastic analysis using only
the reduced distribution. Other points to keep in mind are that, in
some cases, to leading order in $\e$, $Det~\Xi^{(\f)}=0$ (therefore
the distribution function is not independent of the cutoff), and that
the cross-term proportional to $\f_c p_c$ represents a nontrivial
contribution from quantum correlations. In order to discuss these
issues more concretely we now return to the specific cases studied
earlier.

We consider first the massive free scalar field in an exponentially
expanding Universe. At late times, with $\e$ satisfying the condition
(\ref{masscon}), we have,
\bq
f_{cl}(\f_c,p_c)={12\p\over m H_0^{~2} S^3}\exp\left(-{12\p^2\over
H_0^{~2}}\left[\f_c^{~2} - {6\over m^2 S^3}\f_c p_c + {1\over m^2
S^6}\left({9 H_0^{~2}\over m^2}+1\right)p_c^{~2}\right]\right).
\label{csthree}
\eq
Clearly this distribution is not stationary because of the dependence
on $S(\h)$. On the other hand, the corresponding reduced distribution
has the ``equilibrium'' form
\ba
f_{r}(\f_c)&=&{2m\over H_0^{~2}}\sqrt{{\p\over
3}}\exp\left(-{4m^2\p^2\over 3H_0^{~4}}\f_c^{~2}\right)
\label{csfour}\\
&=&m\left({V_{H}\over 2\p
T_{GH}}\right)^{1/2}\exp\left(-\b_{GH}E_H(\f_c)\right) \label{csfive}
\ea
where $\b^{-1}_{GH}\equiv T_{GH}\equiv H_0/2\p$ is the Gibbons-Hawking
temperature of de Sitter space \cite{kn:gibhaw}, $V_H=4\p H_0^{-3}/3$
is the three-volume within the Hubble radius, and $E_H(\f_c)\equiv
V(\f_c)V_H$ is the energy of the scalar field within that volume (the
kinetic energy is not important if the field is in the ``slow-roll''
regime). The thermodynamic interpretation of the stochastic formalism
\cite{kn:tgp} was suggested by the striking Boltzmann-like nature of
(\ref{csfive}). However, the full distribution (\ref{csthree}) does
not seem to encourage such speculation: it is not stationary nor of
the form $e^{-\b H}$ (nor is the reduced distribution $f_r(p_c)$ of
the form $e^{-\b E_{kin}}$).

We observe that transforming to a new variable $v_c=p_c/S^3$ makes the
late time distribution (\ref{csfive}) time independent. However this
transformation is not canonical and does not preserve the linear
entropy. Therefore the distribution $f(\f_c,v_c)$ is not physical. In
any case such a trick fails for the case of power law inflation: there
the phase space distribution cannot be made time independent.

Turning now to the massless case, at late times,
\bq
f_{cl}(\f_c,p_c)={1\over
\e^2S^3}\sqrt{{24\p^2\over H_0^{~7}(t-t_0)}} \exp\left(-{2\p^2\over
H_0^{~2}}\left[{\f_c^{~2}\over H_0(t-t_0)}-{4\f_c
p_c\over\e^2S^3H_0^{~2}(t-t_0)} +
{6p_c^{~2}\over\e^4S^6H_0^{~2}}\right]\right).          \label{cssix}
\eq
The singular nature of this solution as $\e\rightarrow 0$ is apparent
(as is the fact that it is explicitly time dependent). Note, however,
that in this case,
\bq
f_{r}(\f_c)=\sqrt{2\p\over H_0^{~3}(t-t_0)}\exp\left(-{2\p^2\over
H_0^{~3}(t-t_0)}\f_c^{~2}\right)          \label{cseven}
\eq
which is independent of $\e$. The late time reduced distribution
(\ref{cseven}) is a solution of the diffusion equation
\bq
{\pa\over\pa t}f_r(\f_c)={1\over 2}D{\pa^2\over\pa
\f_c^{~2}}f_r(\f_c)                           \label{cseight}
\eq
with $D=H_0^{~3}/4\p^2$. This is suggestive of a (usual, time
independent) random walk interpretation. However, the time dependence
of the terms $\propto p_c^{~2}$ and $\propto \f_cp_c$ in the full
distribution are hard to reconcile with this view.

The case of a massless field in a power law spacetime is treated next.
Here the late time full and reduced distributions are, respectively:
\bq
f_{cl}(\f_c,p_c)={4\p\over\e^2}{\h_0^{~3}\over S^3}\left(t\over
t_0\right)^2\sqrt{3\over
p}\exp\left(-4\p^2\h_0^{~2}\left[{\f_c^{~2}\over p} +
{4\h_0\over\e^2 S^3 p}\left({t\over t_0}\right)\f_c p_c +
{3\h_0^{~2}\over\e^4 S^6}\left({t\over
t_0}\right)^4 p_c^{~2}\right]\right),          \label{csnine}
\eq
\bq
f_{r}(\f_c)=2\h_0\sqrt{\p\over p}\exp\left(-{4\p^2\h_0^{~2}\over
p}\f_c^{~2}\right) .                \label{csten}
\eq
As with the massless field result (\ref{cssix}), the full distribution
is again singular in the limit $\e\rightarrow 0$. Also the reduced
distribution (\ref{csten}) is independent of $\e$, as in the other
cases. Unlike the other two cases however, it does not seem to have
any ``natural'' interpretation.

The late time linear entropies $\s=\int d\f_cdp_c f_{cl}^{~2}$ for the
three cases studied above are respectively:
\ba
\s_{dsm}&=&{6\p\over mH_0^{~2}S^3}={6\p\over
mH_0^{~2}}\hbox{e}^{-3H_0t}, \label{dsm}\\
\s_{ds}&=&{\sqrt{6}\p\over\e^2S^3}{1\over\sqrt{H_0^{~7}(t-t_0)}}=
{\sqrt{6}\p\over\e^2}{\hbox{e}^{-3H_0(t-t_0)}\over
\sqrt{H_0^{~7}(t-t_0)}}, \label{ds}\\
\s_{pl}&=&{6\p\h_0^3\over \e^2\sqrt{3p}S^3}\left({t\over
t_0}\right)^2= {6\p\h_0^3\over \e^2\sqrt{3p}}\left({t\over
t_0}\right)^{3p-2}. \label{pl}
\ea
In all cases $\s$ is approximately proportional to $S^{-3}$. The
possible significance of this result will be discussed later below.

It is by now clear that the late time phase space distributions
obtained here are very difficult to fit into a conventional Brownian
motion picture. In fact, this is a very obvious point and manifest in
our stochastic Hamiltonian (\ref{athirtynine}). In standard Brownian
motion the environment with which the system interacts produces both
dissipative and diffusive effects. The dissipative effects arise from
the back reaction of the environment. Such an effect is absent in
stochastic Hamiltonians of the type (\ref{athirtynine}). In principle,
then, there simply cannot be a fluctuation-dissipation theorem of the
usual sort: this conclusion is manifest in the fact that in no case
are our late time solutions for the distribution function stationary.
(However, this does not mean that there cannot be asymptotically
constant values for some average quantities.)

We recall that the origin of the stochastic noise is simply because
the ``system size'' is changing with time and not because of some
external interaction. It appears that a mistreatment of this key point
has led some authors to claim that quantum decoherence occurs in this
model. That in fact it does not can be explained by the following
direct argument for which the author is indebted to Juan Pablo Paz.

The transition from quantum to classical was studied in the context of
stochastic inflation by Morikawa \cite{kn:mm} and Nambu \cite{kn:yn}
who analyzed the properties of the evolution operator for the reduced
density matrix of the long wavelength modes.  When this propagator is
written in path integral form the effect of the short wavelength modes
appears to be contained in a term that is rather similar to the
Feynman-Vernon influence functional \cite{kn:fv}, $F=\exp i\Gamma$. In
ordinary open systems, the presence of an imaginary part in the
influence action $\Gamma$ produces a tendency towards diagonalization
of the reduced density matrix in a fixed basis. This is known as
decoherence. For stochastic inflation, the imaginary part
of $\Gamma$ was calculated and related to decoherence. However, it is
possible to show that this interpretation is not correct and that
there is no decoherence produced by the coarse-graining of stochastic
inflation. The basic reason is that the time dependent nature of the
coarse-graining prevents us from interpreting the influence functional
in the usual way. In fact the reduced density matrix at a given time can be
written as the product $\prod_{k<k_S(t)}\rho_k$. As the number of modes
present in the system varies with time, the evolution operator
$J(t,t_0)$ has some peculiar properties. It can  be written as a
product of an evolution operator for each mode $J_k(t,t_0)$ where for
$k<k_S(t_0)$ (modes that were already present in the system at
$t=t_0$), the $J_k(t,t_0)$ are ordinary unitary operators while for
$k_S(t_0)<k<k_S(t)$ (modes that enter the system between $t_0$ and $t$)
the evolution operator is simply $J_k(t,t_0)=\rho_k(t)$. If one writes
these operators in path integral form one realizes that there are real
exponential terms simply due to the fact that, if the state of the
field is the vacuum, the reduced density matrix
$\rho_r(\phi_k,\phi_k')$ is a Gaussian. The only effect that the
``influence functional'' has in this case is to generate the above
Gaussian factors. It is clear that this is not related to
decoherence but to the fact that new modes are entering into the
system and that the evolution operator fully contains the reduced
density matrix of the incoming modes.

Another argument put forward for a late time classical limit was that
since the commutator (\ref{thirtytwo}) is $\propto \e^3$, it is ``small'' and
can be ignored. This of course cannot be correct. The reason is that
it is not just a single mode commutator one has to look at but the
total integrated contribution from the initial time to the final time
of interest. This is not a negligible fraction.

Is there a classical limit or not intrinsic to the formalism? The
linear entropy does decrease exceedingly rapidly as shown by
(\ref{dsm})--(\ref{pl}) but it is not clear what this means: we have
just argued against interpreting this sort of decrease as being due to
quantum decoherence. An intuitive basis for this result may be that it
reflects the loss of information inherent in our time dependent
coarse-graining. With the passage of cosmic time, two-point functions are
averaged over ever smaller comoving volumes. The ``smearing'' scale is
set by $k_S^{-1}=\h/\e$ and $\h\rightarrow 0$ as $t\rightarrow\infty$.
Since in our formalism we are tracking only one coarse-grained domain
throughout its history this represents a loss of information with
cosmic time. We may speculate plausibly that the decrease of $\s$ as
$S^{-3}$ supports this viewpoint. However, just because our knowledge
is incomplete is no reason to suppose that the Universe is becoming
more classical!

At the present stage of analysis and understanding it appears unlikely
that the quantum to classical transition in the early Universe can be
explained by the stochastic paradigm. In particular, the treatment of
density perturbations by modeling quantum fluctuations as classical
noise appears to be unjustified.

\newpage
\centerline{\bf V. Conclusion}

This paper's main concern was to model a free field theory in an
inflationary Universe by way of a stochastic quantum
Hamiltonian. It was shown that this approach produced results that
agreed well with those from straightforward quantum field theory.
Furthermore, the role of the length scale parameter $\e$ and of
initial conditions was considered more fully than in previous work.

The quantum phase space distribution used in this paper enables a
consideration of quantum correlations that would otherwise be missed.
As a result we find that the $\e\rightarrow 0$ limit is singular as
far as the distribution function is concerned. This means that a
finite value of $\e$ is essential for the formalism to make sense and
that contrary to previous belief this parameter does not drop out of
the problem. The full phase space distribution also enables a critical
assessment of such issues as the existence of fluctuation-dissipation
relations. We showed that fluctuation-dissipation relations do not
hold (as indicated by the fact that the late time solutions are not
thermal or even stationary). However, at least in de Sitter space, the
reduced distributions for the field variable alone have very
suggestive forms corresponding as they do to a Boltzmann distribution
at the Gibbons-Hawking temperature for the massive field, and to a
``random walk'' distribution for the massless field. No such simple
distribution appears in the case of a power law inflation. The
significance of these results remains unclear at present.

We found that in order to obtain results more or less consistent with
conventional field theoretic calculations quantum correlations could
not be neglected. It was also pointed out that quantum decoherence
does not occur in the stochastic approach. As a consequence of these
two results, the quantum to classical transition in the early Universe
does not seem to be intrinsic to the stochastic approach. Directly
modeling quantum fluctuations by classical noises as a way to study
density perturbations from inflation is therefore a questionable
enterprise.

There are of course many unanswered questions, chief among them is
what happens when interacting fields are considered and back reaction
is included. This we leave to future work. Furthermore, while it is
true that the stochastic model ``works,'' at least to some extent,
we have stressed that it is not free from interpretational problems.
One can only speculate  whether insights gained from this approach
will actually turn out to be valuable when the full quantum field
theoretic computations are eventually done.
\newpage

\centerline{\bf Appendix}

A brief review of conventional field theoretic computations of the
various expectation values of interest will now be given. This will
enable us to check results from the stochastic analysis. To obtain
finite results we will impose an infrared cutoff in momentum space
at $k=\h_0^{-1}$ and an ultraviolet cutoff at $k=\e\h^{-1}$. A
discussion of the reasons for picking these cutoffs will be given at
the end of the Appendix.

We begin with exponential inflation and treat the two cases of a
massless field and of a massive field with a small mass $(m^2\ll
H_0^{~2})$. Considering first the case of a massless field, the
parameter $\n=3/2$ (from (\ref{cfive})), and for the Bunch-Davies
vacuum the mode functions $\f_k\equiv\c_k/S$ are
\bq
\f_k(\h)=\left(\p\over 4\right)^{1/2}H_0\h^{3/2}H_{3/2}^{(2)}(k\h).
\label{Aone}
\eq
Consider first, the equal time expectation value,
\bq
\VEV{\F(\vec{x})\F(\vec{y})}={1\over 2\p^2}\int^{\e\h^{-1}}_{\h_0^{-1}}
dk~k^2{\sin{kR}\over kR}\abs{\f_k(\h)}^2,    \label{Atwo}
\eq
where $R=\abs{\vec{x}-\vec{y}}$. Using the exact form of the Hankel
function (\ref{csixteen}) it is easy to see that
\bq
\abs{\f_k(\h)}^2={H_0^{~2}\over 2k^3}\left[1+(k\h)^2\right],
\label{Athree}
\eq
hence the integral in (\ref{Atwo}) is infrared divergent and an
infrared cutoff is necessary. With $kR\ll 1$, we find
\bq
\VEV{\F^2}={H_0^{~2}\over
4\p^2}\left[\ln\left({\e\h_0\over\h}\right)+{1\over 2}\e^2-{1\over
2}\left({\h\over\h_0}\right)^2\right].  \label{Afour}
\eq
Notice that since $\e\h^{-1}>\h_0^{-1}$, $\VEV{\F^2}$ as computed
above is strictly positive. To write the result in terms of the cosmic
time, we note that $\h_0\h^{-1}=\exp{H_0(t-t_0)}$, in which case
\bq
\VEV{\F^2}={H_0^{~3}\over 4\p^2}(t-t_0)+{H_0^{~2}\over
4\p^2}\left(\ln\e+{1\over 2}\e^2\right)-{H_0^{~2}\over
8\p^2}\hbox{e}^{-2H_0(t-t_0)}.            \label{Afive}
\eq
The last term vanishes at late times and the second term is an
irrelevant constant absorbed in the infrared cutoff. The first term
gives the usual result \cite{kn:destab}. Notice that this term is
independent of $\e$; any potential dependence is lost in the infrared
cutoff.

Now we turn to the quantity
\bq
{1\over 2}\VEV{\F\dot{\F}+\dot{\F}\F}={1\over
2\p^2}\int^{\e\h^{-1}}_{\h_0^{-1}}dk~k^2
Re\left(\f_k(\h)\dot{\f}_k^*(\h)\right)         \label{Asix}
\eq
where we have already set $kR\ll 1$. From (\ref{Aone}),
\bq
\dot{\f}_k=\left({\p\over 4}\right)^{1/2}H_0k\h^{3/2}
H_{1/2}^{(2)}(k\h), \label{Aseven}
\eq
and it is easy to compute that
\bq
{1\over 2}\VEV{\F\dot{\F}+\dot{\F}\F}={H_0^{~2}\over
8\p^2}\h^{-1}\left(\e^2 - \left({\h\over\h_0}\right)^{~2}\right).
\label{Aeight}
\eq
If we return to the cosmic time, then with a prime denoting
differentiation with respect to $t$,
\bq
{1\over 2}\VEV{\F\F'+\F'\F}={H_0^{~3}\over 8\p^2}\left(\e^2 -
\hbox{e}^{-2H_0(t-t_0)}\right).              \label{Anine}
\eq
The last term vanishes at late times and the first term is a constant
that is strongly dependent on the upper cutoff. (Even if $\e\ll 1$,
consistency requires $\e^2\gg \left({\h/\h_0}\right)^2$ and the first
term always dominates.) The late time answer being a strong function
of the cutoff is simply a consequence of the extra multiplicative
factor of $k$ in $\dot{\f}_k$ which not only renders the integral in
(\ref{Asix}) infrared finite but also shifts the dominant
contribution from the integrand towards the upper cutoff.

To compute $\VEV{\dot\F^2}$ we first use (\ref{Aseven}) to show that
\bq
\abs{\dot{\f}_k(\h)}^2={1\over 2}H_0^{~2}\h^2k.      \label{Aten}
\eq
With $kR\ll 1$, it is now easy to find
\ba
\VEV{\dot{\F}^2}&=&{1\over 2\p^2}\int_{\h_0^{-1}}^{\e\h^{-1}}
dk~k^2\abs{\dot{\f}_k(\h)}^2\nonumber\\
&=&{H_0^{~2}\over 16\p^2}\h^{-2}\left[\e^4-\left({\h\over
\h_0}\right)^4\right],    \label{Aeleven}
\ea
or, in terms of the cosmic time,
\bq
\VEV{\F'^2}={H_0^{~4}\over
16\p^2}\left[\e^4-\hbox{e}^{-4H_0(t-t_0)}\right].   \label{Atwelve}
\eq
The late time value is again a strongly cutoff dependent constant.

Similar calculations will now be performed for the case of a small
mass, i.e., for $m^2\ll H_0^{~2}$. Disregarding an irrelevant phase term
for real $\n$, the mode functions are now
\bq
\f_k(\h)=\left(\p\over 4\right)^{1/2}H_0\h^{3/2}H_{\n}^{(2)}(k\h)
\label{Athirteen}
\eq
where, for a small mass,
\bq
\n\simeq{3\over 2}-{m^2\over 3H_0^{~2}}.    \label{Afourteen}
\eq
Since $k\h<1$ over the range of integration $\h_0^{-1}<k<\e\h^{-1}$,
we may approximate the Hankel functions by
\bq
H_{\n}^{(2)}(z){\stackrel{z\rightarrow 0}{\longrightarrow}}-
{i\over\p}\G(\n)\left({z\over 2}\right)^{-\n},    \label{Afifteen}
\eq
and using (\ref{Atwo}) compute,
\ba
\VEV{\F^2}&\simeq& {3H_0^{~4}\over 8\p^2m^2}\left[\e^{2m^2/3H_0^{~2}}-
\left({\h\over\h_0}\right)^{2m^2/3H_0^{~2}}\right] \label{Asixteen}\\
&=&{3H_0^{~4}\over 8\p^2m^2}\left[\e^{2m^2/3H_0^{~2}}-
\exp\left(-{2m^2\over
3H_0}(t-t_0)\right)\right].\label{Aseventeen}
\ea
As long as $\e\gg\exp(-3H_0^{~2}/2m^2)$ (i.e., $\e$ cannot be
arbitrarily small),
\bq
\e^{2m^2/3H_0^{~2}}\simeq 1 + {2m^2\over 3H_0^{~2}}\ln{\e} + \cdots ,
\label{Aeighteen}
\eq
and at late times
\bq
\VEV{\F^2}={3H_0^{~4}\over 8\p^2m^2},         \label{Anineteen}
\eq
a well known result \cite{kn:destab}.

The calculational strategy used above can also be implemented to find
that
\bq
{1\over 2}\VEV{\F\dot{\F}+\dot{\F}\F}\simeq {H_0^{~2}\over
8\p^2}\h^{-1}\left[\e^{2m^2/3H_0^{~2}}+\e^2-\left({\h\over\h_0}
\right)^{2m^2/3H_0^{~2}}\left(1+\left({\h\over\h_0}\right)^2
\right)\right],               \label{Atwenty}\\
\eq
and that,
\bq
{1\over 2}\VEV{\F\F'+\F'\F}\simeq {H_0^{~3}\over 8\p^2}
\left[\e^{2m^2/3H_0^{~2}}+\e^2-\exp\left(-{2m^2\over
3H_0^{~2}}(t-t_0)\right)\left(1+\hbox{e}^{-2H_0(t-t_0)}
\right)\right].                 \label{Atwentyone}
\eq
Setting $m^2=0$ in (\ref{Atwentyone}) we recover the previous result
(\ref{Anine}) for a massless scalar field. The late time limit for
$\e\gg\exp(-3H_0^{~2}/2m^2)$ is
\bq
{1\over 2}\VEV{\F\F'+\F'\F}={H_0^{~3}\over 8\p^2}\left(1+\e^2\right).
\label{Atwentytwo}
\eq
Unlike the massless case, here the result is essentially $\e$
independent provided $\e$ is small compared to one. However, unlike
the situation for $\VEV{\F^2}$, the late time value does depend on
whether $\e$ is small or of order unity.

The expectation value $\VEV{\F'^2}$ can be found in exactly the
same way:
\ba
\VEV{\F'^2}&\simeq&{H_0^{~4}\over 4\p^2}\left[{m^2\over
6H_0^{~2}}\left\{\e^{2m^2/3H_0^{~2}}-\exp\left(-{2m^2\over
3H_0}(t-t_0)\right)\right\}\right. \nonumber\\
&& + \left.{1\over 4}\left\{\e^4 -\hbox{e}^{-4H_0(t-t_0)}\right\}
+ {2m^2\over 3H_0^{~2}}\left\{\e^2-\hbox{e}^{-2H_0(t-t_0)}
\right\}\right].              \label{Atwentythree}
\ea
If we assume $\e$ to be much smaller than unity, then at late times we
obtain to leading order an $\e$ independent result,
\bq
\VEV{\F'^2}\simeq {m^2H_0^{~2}\over 24\p^2}. \label{Atwentyfour}
\eq
On the other hand if we set $\e$ to be one, then the late time limit
becomes
\bq
\VEV{\F'^2}\simeq {H_0^{~4}\over 16\p^2} \label{Atwentyfive}
\eq
which is completely different from (\ref{Atwentyfour}). It is easy to
verify that setting $m^2=0$ in (\ref{Atwentythree}) reproduces the
answer (\ref{Atwelve}) for the massless case. Therefore for $\e\sim
1$, the massless and massive cases give the same result (but not when
$\e$ is small).

We turn now to power law inflation and consider the case of a massless
field. Recall that the parameter $\n$ is now given by
\bq
\n={1-3p\over 2(1-p)}       \label{Atwentysix}
\eq
and that the adiabatic vacuum modes are
\bq
\f_k(\h)=\left({\p\h_0\over 4}\right)^{1/2}\left({\h\over
\h_0}\right)^{\n}H_{\n}^{(2)}(k\h).   \label{Atwentyseven}
\eq
The expectation value $\VEV{\F^2}$ is still given by (\ref{Atwo}) and
we can still use the approximation (\ref{Afifteen}) for the Hankel
function. If we set $\n=3/2$ then the de Sitter results for a massless
field are recovered. For $\n\neq 3/2$, we have
\ba
\VEV{\F^2}&=&{2^{2\n-3}\over \p^3}\G(\n)^2\h_0^{1-2\n}
\int_{\h_0^{-1}}^{\e\h^{-1}}dk~k^{2(1-\n)}.
\label{Atwentyeight}\nonumber\\
&=&{2^{2\n-3}\G(\n)^2\over \p^3(2\n-3)}\h_0^{-2}\left[1-
\left({\e\h_0\over \h}\right)^{3-2\n}\right]
\label{Atwentynine}\nonumber\\
&=&{2^{2\n-3}\G(\n)^2\over
\p^3(2\n-3)}\h_0^{-2}\left[1-\e^{2/(1-p)}\left({t_0\over
t}\right)^2\right] \label{Athirty}
\ea
reproducing the result of \cite{kn:despl}. (As long as the power law
$p>1$, the parameter $\n>3/2$, and it follows that the integral in
(\ref{Atwentyeight}) is infrared divergent. The lower cutoff is
necessary to get a finite answer.) We see that $\VEV{\F^2}$ starting
from some initial value rises to a constant. When the power law $p\gg
1$, (\ref{Atwentysix}) implies that $\n\simeq 3/2 +1/p$, and
\bq
\VEV{\F^2}\simeq {p\over 8\p^2}\h_0^{-2}\left[1-\e^{-2/p}
\left({t_0\over t}\right)^2\right].        \label{Athirtyone}
\eq
If $\e>>\hbox{e}^{-p/2}$, then
\bq
\e^{-2/p}\simeq 1 - {2\over p}\ln{\e} + \cdots \label{Athirtytwo}
\eq
and (\ref{Athirtyone}) is essentially cutoff independent not only as
to the late time constant value but also as to how this value is
approached. The situation is different when $p$ is not large. For
example, if we consider $p=2$ corresponding to $\n=5/2$,
\bq
\VEV{\F^2}={9\over 8\p^2}\h_0^{-2}\left[1-\e^{-2}\left({t_0\over
t}\right)^2\right].               \label{Athirtythree}
\eq
This time while the asymptotic value of $\VEV{\F^2}$ is indeed
cutoff independent, the approach to it is a strong function of
$\e$.

We proceed now to evaluate $\VEV{\F\dot{\F}+\dot{\F}\F}/2$. The
usual procedure, beginning from (\ref{Asix}), yields
\bq
{1\over 2}\VEV{\F\dot{\F}+\dot{\F}\F}\simeq{2^{2\n-4}\over
\p^3}\G(\n)\G(\n-1)\h_0^{1-2\n}\h\int_{\h_0^{-1}}^{\e\h^{-1}}
dk~k^{4-2\n}.             \label{Athirtyfour}
\eq
The integral in (\ref{Athirtyfour}) is infrared divergent for $\n\geq
5/2$ (i.e., $1<p\leq 2$). For $\n=5/2$,
\bq
{1\over 2}\VEV{\F\dot{\F}+\dot{\F}\F}\simeq {3\over
4\p^2}\h_0^{-4}\h\ln{\left({\e\h_0\over \h}\right)} \label{Athirtyfive}
\eq
which vanishes at late times. In terms of the cosmic time,
\bq
{1\over 2}\VEV{\F\F'+\F'\F}\simeq {3\over
4\p^2}\h_0^{-3}\left({t_0\over t}\right)^3\ln{\left({\e t\over
t_0}\right)}.          \label{Athirtysix}
\eq
After an initial period of growth, at late times this expectation
value vanishes.

When $p\neq 2$, we obtain from (\ref{Athirtyfour})
\bq
{1\over 2}\VEV{\F\F'+\F'\F}\simeq {2^{2\n-1}\over
8\p^3(5-2\n)}\G(\n)\G(\n-1) \h_0^{-3}\left({t_0\over t}\right)^3
\left[\e^{5-2\n}-\left({t_0\over t}\right)^{2p-4}\right]. \label{Athirtyseven}
\eq
It is trivial to check that this quantity is always positive. At late
times it vanishes but the dependence on $\e$ is a function of the
power law. For large $p$, (\ref{Athirtyseven}) becomes
\ba
{1\over 2}\VEV{\F\F'+\F'\F}&\simeq&{\h_0^{-3}\over
8\p^2}\left({\h\over\h_0}\right)^{3/p}\left[\e^2 -
\left({\h\over\h_0}\right)^{2-2/p}\right]  \label{Athirtyeight}\\
&\simeq&{\e^2\over 8\p^2}\h_0^{-3}\left({t_0\over
t}\right)^3.                                   \label{Athirtynine}
\ea
If $p\rightarrow\infty$, (\ref{Athirtyeight}) reproduces the de Sitter
result (\ref{Anine}) for a massless field (with $H_0=\h_0^{-1}$).

Finally we compute $\VEV{\dot{\F}^2}$. The standard calculation yields
\bq
\VEV{\dot{\F}^2}\simeq {\h_0^{1-2\n}\over 8\p^3}2^{2(\n-1)}\G(\n-1)^2
\h^2\int_{\h_0^{-1}}^{\e\h^{-1}}dk~k^{6-2\n}.   \label{Athirtyeighta}
\eq
The $k$ integral is infrared divergent provided $\n\geq 7/2$ (i.e.,
$1<p\leq 3/2$). For the special case $\n=7/2$ corresponding to $p=3/2$,
we have
\bq
\VEV{\dot{\F}^2}\simeq {9\over
4\p^2}\h_0^{-6}\h^2\ln{\left({\e\h_0\over \h}\right)} \label{Athirtyninea}
\eq
or,
\bq
\VEV{\F'^2}\simeq {9\over 8\p^2}\h_0^{-4}\left({t_0\over
t}\right)^4\ln{\left({\e^2t\over t_0}\right)}.      \label{Aforty}
\eq
At late times this expectation value vanishes.

In the general case ($\n\neq 7/2$) we find
\bq
\VEV{\F'^2}\simeq {2^{2(\n-1)}\h_0^{-4}\over 8\p^3(7-2\n)}\G(\n-1)^2
\left({t_0\over t}\right)^4\left[\e^{7-2\n}-\left({t_0\over
t}\right)^{2(2p-3)}\right]              \label{Afortyone}
\eq
which also vanishes at late times. When $p\gg 1$, (\ref{Afortyone})
gives to leading order,
\ba
\VEV{\F'^2}&\simeq&{\h_0^{-4}\over 16\p^2}\left({\h\over
\h_0}\right)^{4/p}\left[\e^4 -
\left({\h\over\h_0}\right)^{4-2/p}\right]          \label{Afortytwo}\\
&\simeq&{\e^4\over 16\p^2}\h_0^{-4}\left({t_0\over t}\right)^4.
\label{Afortythree}
\ea
The limit $p\rightarrow\infty$ taken in (\ref{Afortytwo}) gives back
the de Sitter result (\ref{Atwelve}) for a massless field.

We now explain the origin of the cutoffs in the momentum integrals.
To prevent infrared divergences we follow the strategy of Ford and
Parker \cite{kn:fp} by assuming that for $\h_0>\h$ the Universe is
radiation dominated and that the quantum state is the conformal
vacuum. Matching the field modes and their time derivatives at
$\h=\h_0$, one finds
\bq
\abs{C_1(k)-C_2(k)}^2={1\over 1+\left({2^{2\n}\over
2\p}\right)\left(k\h_0\right)^{1-2\n}\G(\n)^2}.      \label{Afortyfour}
\eq
Since the upper cutoff forces $k\h<1$, we can use the small argument
form (\ref{Afifteen}) of the Hankel function and compute
\bq
\abs{\f_k}^2\simeq {2^{2\n}\over
4\p}\G(\n)^2\h_0^{1-2\n}k^{-2\n}\abs{C_1(k)-C_2(k)}^2.  \label{Afortyfive}
\eq
We are now in a position to compute the pre-inflationary contributions
to the various expectation values of interest. First, consider
\ba
\VEV{\F^2}_{PI}&=&{1\over 2\p^2}\int_0^{\h_0^{-1}}dk~k^2\abs{\f_k}^2
\nonumber\\
&\simeq&{1\over 4\p^2}\int_0^{\h_0^{-1}}dk~k \nonumber\\
&=&{\h_0^{-2}\over 8\p^2},      \label{Afortysix}
\ea
which is a constant {\em independent} of $\n$. In a similar fashion we
can evaluate
\bq
{1\over 2}\VEV{\F\F'+\F'\F}_{PI} \simeq {\h_0^{-3}\over
32\p^2(\n-1)}\left({\h\over\h_0}\right)^{\n+1/2}    \label{Afortyseven}
\eq
and
\bq
\VEV{\F'^2}_{PI} \simeq {\h_0^{-4}\over
96\p^2(\n-1)^2}\left({\h\over\h_0}\right)^{2\n+1}.  \label{Afortyeight}
\eq

If we restrict attention to de Sitter space, then
(\ref{Afortysix})--(\ref{Afortyeight}) specialize to
\ba
\VEV{\F^2}_{PI}&\simeq&{H_0^{~2}\over 8\p^2}, \label{Afortynine}\\
{1\over 2}\VEV{\F\F'+\F'\F}_{PI}&\simeq&{H_0^{~3}\over
16\p^2}\hbox{e}^{-2H_0(t-t_0)}, \label{Afifty}\\
\VEV{\F'^2}_{PI}&\simeq&{H_0^{~4}\over 24\p^2}\hbox{e}^{-4H_0(t-t_0)}.
\label{Afiftyone}
\ea
Comparison with (\ref{Afive}), (\ref{Anine}), and (\ref{Atwelve})
shows that the late time results are unaffected: the dominant
contribution to these expectation values comes from the inflationary
sector.

In the case of a power law inflation, we find
\ba
\VEV{\F^2}_{PI}&\simeq&{\h_0^{-2}\over 8\p^2}, \label{Afiftytwo}\\
{1\over 2}\VEV{\F\F'+\F'\F}_{PI}&\simeq&{\h_0^{-3}\over
16\p^2}\left({p+1\over p-1}\right)\left({t_0\over t}\right)^{2p-1},
\label{Afiftythree}\\
\VEV{\F'^2}_{PI}&\simeq&{\h_0^{-4}\over 24\p^2}\left({p-1\over
p+1}\right)^2\left({t_0\over t}\right)^{2(2p-1)}. \label{Afiftyfour}
\ea
The late time constant value of $\VEV{\F^2}$ does get shifted due to
(\ref{Afiftytwo}) but for large $p$ this shift is negligible as
comparison with (\ref{Athirtyone}) shows. Even for relatively small
values of $p$, this term is relatively unimportant (compare with
(\ref{Athirtythree}) for $p=2$). From (\ref{Athirtynine}) and
(\ref{Afortythree}) we know that for $p\gg 1$, the contributions from
the inflationary sector to $\VEV{\F\F'+\F'\F}/2$ and $\VEV{\F'^2}$ fall
off as $(t_0/t)^3$ and $(t_0/t)^4$ respectively. These fall-offs are
much slower than those given by (\ref{Afiftythree}) and
(\ref{Afiftyfour}): again the pre-inflationary contributions are
insignificant. It is only for weak power law expansions ($p \sim 1$)
that this sector is of any significance.

The rationale for the upper cutoff is simple. In quantum field theory
in curved spacetime nontrivial ultraviolet divergences can arise
because of the spacetime curvature. In principle one has to apply an
appropriate regularization scheme (point-splitting, for example)
followed by an ultraviolet subtraction. Following this more
sophisticated procedure one obtains terms proportional to the
curvature in $\VEV{\F^2}$. However, these terms either vanish at late
times (power law inflation) or are constants which are small compared
to the contribution from the infrared sector (exponential inflation).
In this sense our procedure is justified. (The fact that $k\h\sim 1$
separates the low and high frequency sectors is due to the following
behavior of the Hankel functions: oscillatory for $k\h>>1$, and power
law for $k\h<<1$.)

\centerline{\bf Acknowledgments}
It is a pleasure to thank Larry Ford, Bei-lok Hu, Henry Kandrup, Milan
Miji\'c, Juan Pablo Paz, Bill Unruh, Wojciech Zurek and Robert Zwanzig
for helpful discussions. This research was supported by the Natural
Sciences and Engineering Research Council of Canada, the Canadian
Institute for Advanced Research, the Canadian Institute for Theoretical
Astrophysics, and the United States Department of Energy.

\end{document}